\documentclass[5p,times]{elsarticle}

\usepackage[numbers]{natbib}
\usepackage{float}
\usepackage{orcidlink}
\usepackage{subfig}
\usepackage{graphicx}
\usepackage{url}
\usepackage{xcolor}
\usepackage{stfloats}
\usepackage{tcolorbox}
\usepackage{amsmath}
\usepackage{enumitem}
\usepackage{amsfonts}
\usepackage{multirow}
\usepackage{booktabs}
\usepackage{listings}
\usepackage{pifont}
\usepackage{xcolor}
\usepackage{fvextra}
\usepackage{makecell}
\usepackage{nicematrix}
\usepackage{bm}
\usepackage{threeparttable}

\definecolor{verylightgray}{rgb}{.97,.97,.97}
\DefineVerbatimEnvironment{MyVerbatim}{Verbatim}
{breaklines=true, formatcom=\color{blue}}

\journal{Information and Software Technology}

\begin{document}

\begin{frontmatter}
	
\title{Evaluating and Improving LLM-based Competitive Program Generation}

\author[NTU]{Minnan Wei\orcidlink{0009-0007-5479-5784}}
\ead{minnanvvei@gmail.com}

\author[NTU]{Ziming Li\orcidlink{0009-0008-6924-8510}}
\ead{dzycd53@gmail.com}

\author[NTU]{Xiang Chen\orcidlink{0000-0002-1180-3891}\corref{mycorrespondingauthor}}
\cortext[mycorrespondingauthor]{Corresponding author}
\ead{xchencs@ntu.edu.cn}

\author[NTU]{Menglin Zheng\orcidlink{0009-0006-3197-8863}}
\ead{2230110478@stmail.ntu.edu.cn}

\author[NTU]{Ziyan Qu\orcidlink{0009-0006-7446-1207}}
\ead{quziyanu@gmail.com}

\author[NTU]{Cheng Yu\orcidlink{0009-0007-3244-166X}}
\ead{zydy1227@gmail.com}

\author[NTU]{Siyu Chen\orcidlink{0009-0006-8951-4922}}
\ead{chensiyu043@gmail.com}

\author[NTU]{Xiaolin Ju\orcidlink{0000-0003-2579-5359}}
\ead{ju.xl@ntu.edu.cn}

	\address[NTU]{School of  Artificial Intelligence and Computer Science, Nantong University, Nantong, China}

\begin{abstract}

\textbf{Context:} Due to the demand for strong algorithmic reasoning, complex logic implementation, and strict adherence to input/output formats and resource constraints, competitive programming generation by large language models (LLMs) is considered the most challenging problem in current LLM-based code generation.
However, previous studies often evaluate LLMs using simple prompts and benchmark datasets prone to data leakage. Moreover, prior work has limited consideration of the diversity in algorithm types and difficulty levels.

\textbf{Objective:} In this study, we aim to evaluate and improve LLMs in solving real-world competitive programming problems.  

\textbf{Methods:} We initially collect 117 problems from nine regional ICPC/CCPC contests held in 2024 and design four filtering criteria to construct a curated benchmark consisting of 80 problems.
Leveraging DeepSeek-R1 as the LLM, we evaluate its competitive program generation capabilities through the online judge (OJ) platforms, guided by a carefully designed basic prompt.
For incorrect submissions, we construct a fine-grained error taxonomy and then propose a targeted improvement framework by combining a multi-turn dialogue-based repair phase and an information-augmented regeneration phase. 

\textbf{Results:} Experimental results show that only 5 out of 80 problems are fully accepted when using basic prompts.
For the unsolved problems, we construct the error taxonomy, including general errors (such as design, boundary, condition, data type, syntax, and input/output errors) and specialized errors (such as those in mathematical problems, greedy algorithms, and graph theories).
After applying our proposed improvement strategies, we substantially increased the number of correct solutions, with 46 out of 80 problems successfully accepted.

\textbf{Conclusion:} Our study highlights the current limitations of LLM-based competitive program generation and outlines promising directions for improving the performance.

\end{abstract}


\end{frontmatter}

\section{Introduction}
\label{sec:introduction}

\textbf{Background.} The primary goal of LLM-based code generation~\cite{jiang2024survey} is to automate software development by generating functional, efficient, and context-aware code from natural language descriptions or partial inputs. This technology improves developer productivity by reducing manual coding effort and accelerates software prototyping and maintenance.
Among various code generation tasks, competitive program generation is regarded as the most challenging task. Unlike traditional software development tasks, competitive programming requires LLMs to demonstrate precise algorithmic reasoning. The LLMs should fully understand the problem description and independently design and implement efficient algorithms. These problems frequently involve complex logic and extensive edge case handling, further increasing the difficulty. Additionally, solutions should strictly conform to specified input/output formats and adhere to given performance constraints, such as time and memory limits, posing a substantial challenge to the LLM’s ability to generate correct and optimized code.

\textbf{Research Motivation.} Although an increasing number of studies~\cite{zhang2022automated,mhossain2025ll,li2023competition,lu2024magicoder} have focused on leveraging LLMs for competitive program generation, current research still suffers from the following limitations.

First, existing research on LLM-based competitive program generation often lacks task-specific prompt design. Without prompts tailored to the unique requirements of competitive programming, the generated solutions frequently fall short in terms of correctness, efficiency, or adherence to problem constraints~\cite{wang2024enhancing,ridnik2024code}.

Second, existing datasets for competitive program generation often suffer from the issue of data leakage~\cite{wang2025vericontaminated}, as problems or code snippets may have appeared during the pre-training phase of LLMs~\cite{matton2024leakage}. This issue is particularly evident in widely used benchmarks such as HumanEval~\cite{chen2021evaluating} and MBPP~\cite{austin2021program}, which include problems collected from platforms like Codeforces and LeetCode~\cite{riddell2024quantifying}. As a result, it becomes challenging to determine whether LLMs are genuinely solving problems or simply recalling memorized data.

Third, existing datasets rarely incorporate problems from real-world competitive programming contests, resulting in a limited diversity of problem types and difficulty levels~\cite{huang2023competition}. Therefore, evaluations of previous studies may fail to comprehensively assess how LLMs perform across the wide range of reasoning and implementation challenges presented by different difficulty levels and problem types.

\textbf{Benchmark.} To fill this gap, we collected real-world competitive programming problems from five Asian regional contests of the International Collegiate Programming Contest (ICPC) and four regional contests of the China Collegiate Programming Contest (CCPC) held in 2024, thereby helping to mitigate the data leakage issue. The initial pool consists of 117 problems. After applying four carefully designed filtering criteria (such as excluding ``gold-level” problems, geometry problems with visual images, and problem descriptions with difficult content), we ultimately selected 80 problems that better align with the current capabilities of LLMs. In summary, our constructed benchmark contains 13 warm-up problems, 36 bronze-level problems, and 31 silver-level problems, covering a diverse range of algorithm types and difficulty levels.

\begin{figure*}[htbp] 
    \centering 
    \includegraphics[width=0.95\textwidth]{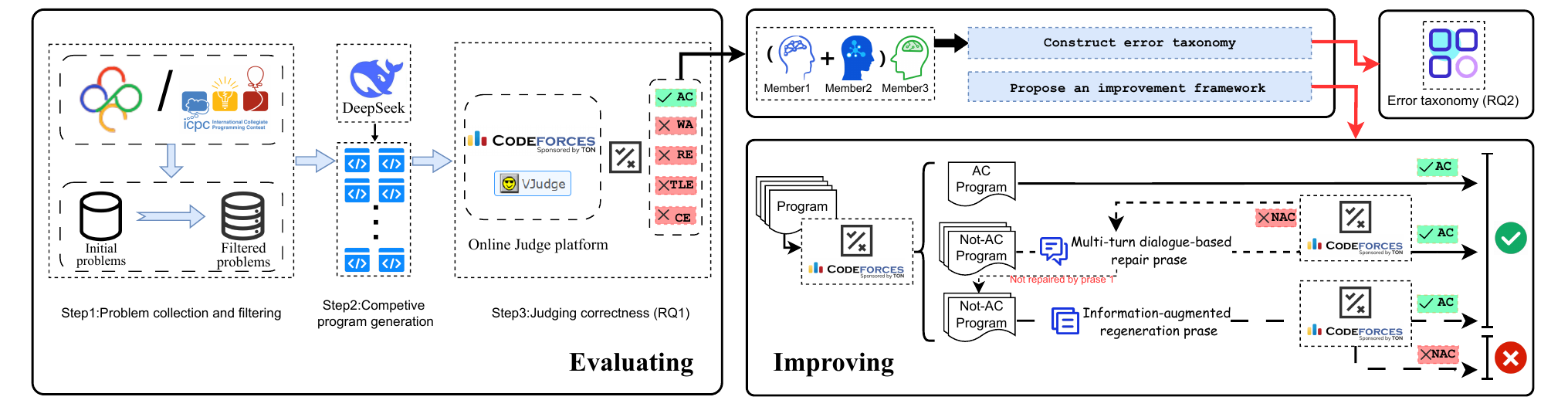}
\caption{Our empirical study methodology for evaluating and improving LLM-based competitive program generation} 
    \label{Fig:Methodologyoverview1} 
\end{figure*}

\textbf{Methodology.} Our empirical study methodology is illustrated in Figure~\ref{Fig:Methodologyoverview1}. Specifically, we first generate initial solutions using our designed basic prompt tailored for the competitive algorithm generation task, leveraging DeepSeek-R1 as the LLM. The correctness of the generated programs is evaluated on the VJudge\footnote{\url{https://vjudge.net/}} and Codeforces\footnote{\url{https://codeforces.com/submissions/xxx}} online judge (OJ) platforms. The results are categorized as Accept (AC), Wrong Answer (WA), Runtime Error (RE), Time Limit Exceeded (TLE), and Compile Error (CE). We then conduct error analysis through LLM-assisted detection and taxonomy-based classification. Finally, we propose a targeted improvement framework based on multi-turn dialogue and information-augmented strategies.

\textbf{LLM Evaluation.} To evaluate the performance of LLM for competitive program generation, we want to answer the following two research questions.

\textbf{RQ1: Given our designed basic prompt, how correct is the competitive program generated by LLM?}

Experimental results show that, under basic prompting, the LLM DeepSeek-R1 fully solves only 5 out of 80 problems. The failures consist of 63 cases of WA, 5 cases of TLE, and 7 cases of CE. In particular, the LLM performs relatively better on warm-up and single-algorithm problems, it struggles significantly with more challenging tasks, such as bronze- and silver-level problems and those requiring multiple algorithmic techniques. These findings highlight the limitations of current LLMs in generating correct and efficient solutions for the competitive program generation task when relying on basic prompts.

\textbf{RQ2: What types of errors commonly occur in incorrect programs generated by LLM?}

We first constructed a comprehensive hierarchical error taxonomy to categorize the common failures of LLM-generated programs systematically.  Based on this taxonomy, in the general error category, error analysis reveals that design-related errors (28.6\%) and boundary-related Errors (15.5\%) are the predominant issues, followed by condition-related errors, data type errors, syntax errors, and input/output errors.  Typical subcategories include incorrect algorithm (12.4\%), misunderstanding problem requirements (10.6\%), and Incorrect Handling of edge cases or input boundaries (8.7\%). In the algorithm-specific error category, we summarize several types of common errors that occur in specific algorithm types, such as mathematical problems and greedy algorithms.

\textbf{LLM Improvement.} To improve LLMs for competitive program generation, we design an improvement framework that combines multi-turn dialogue repair and information-augmented regeneration to enhance program correctness systematically. The framework consists of three phases: \textbf{Phase 1} performs error identification and taxonomy classification to guide subsequent repair; \textbf{Phase 2} applies targeted, multi-turn dialogue prompts to iteratively revise the faulty code; and \textbf{Phase 3} conducts regeneration using structured, problem-specific information when earlier fixes fail. This staged design ensures both precision in fixing localized issues and completeness in regenerating complete solutions. 

\textbf{Improvement Results:} After applying our designed improvement framework, the number of AC solutions increased from 5 to 46 out of 80 problems. Specifically, 33 WA cases, 2 TLE cases, and 6 CE cases were successfully accepted.

\textbf{Impact of Our Study:} Our study systematically evaluates the limitations of the current generation of competitive programs based on LLM using a constructed benchmark gathered from recent real-world algorithmic contests.  Based on our analysis, we propose potential strategies to improve LLM performance. Promising results can provide practical insights to guide future research in this task.

The main contributions of our study can be summarized as follows:

\begin{itemize}
\item \textbf{Challenging Benchmark.} We construct a challenging benchmark for competitive program generation using 80 problems from the 2024 ICPC and CCPC regional contests. Compared to existing datasets, our benchmark helps mitigate the data leakage issue. Furthermore, it can provide broad coverage across diverse problem types and difficulty levels.

\item \textbf{Initial Evaluation.} For the competitive program generation task, we design task-specific basic prompts. Based on the DeepSeek-R1 LLM, we evaluate its performance on our challenging benchmark. The results show that the generation capability is still limited. To better understand the causes of failure, we construct a hierarchical error taxonomy and conduct a detailed analysis for each category.

\item \textbf{Improvement.} Based on our error analysis, we propose a targeted improvement framework that combines multi-turn dialogue repair with information-augmented regeneration. Evaluation results show a substantial performance improvement, with the number of AC solutions increasing from 5 to 46 out of 80 problems, demonstrating the practical effectiveness of our proposed strategies.
\end{itemize}

\textbf{Open science.} 
To facilitate reproducibility and encourage other researchers to follow our study, we share the benchmark, scripts, and detailed results on GitHub\footnote{\url{https://github.com/minnanWei/LLMs-Competitive-Program-Generation.git}}.

 \textbf{Paper Organization.}
The remainder of this paper is organized as follows. Section~\ref{sec:introduction} provides an overview of the study background and motivation. Section~\ref{sec:experimentalsetupofLLMevaluation} describes the benchmark construction and experimental setup. Section~\ref{sec:researchquestions} formulates the key research questions guiding our study. Section~\ref{sec:resultsofLLMevaluation} presents the empirical evaluation results and error analysis. Section~\ref{sec:LLMimprovement} proposes improvement strategies based on the identified errors. Finally, Section~\ref{sec:conclusion} summarizes the key findings and discusses directions for future work.

\section{Experimental Setup of LLM Evaluation}
\label{sec:experimentalsetupofLLMevaluation}

In this section, we describe the experimental setup used to evaluate LLM performance on competitive program generation. Specifically, we first show the construction process of our benchmark. Then, we introduce the LLM used in our experiments. Finally, we show the details of the basic prompt tailored for this task.

\subsection{Benchmark Construction}

In this subsection, we first present the competitive contests used to collect competitive programming problems. Then, we introduce our filtering criteria for selecting problems. Finally, we categorize the selected problems based on their types and difficulty levels.

\subsubsection{Competitive Problem Collection}

To ensure diversity in problem types and difficulty levels, and to mitigate the potential data leakage issue, we selected regional competitive contests from ICPC and CCPC held in 2024.

For the ICPC, we select the following five Asian regional contests and provide their corresponding URLs:
\begin{itemize} 
    \item \textbf{ICPC-CD.} ICPC Asia Chengdu Regional Contest\footnote{\url{https://codeforces.com/gym/105486}}
    \item \textbf{ICPC-NJ.} ICPC Asia Nanjing Regional Contest\footnote{\url{https://codeforces.com/gym/105484}}
    \item \textbf{ICPC-SY.} ICPC Asia Shenyang Regional Contest\footnote{\url{https://codeforces.com/gym/105578}}.
    \item \textbf{ICPC-KMR.} ICPC Asia Kunming Regional Contest\footnote{\url{https://codeforces.com/gym/105588}}.
    \item \textbf{ICPC-KMI.} ICPC Kunming Invitational Contest\footnote{\url{https://codeforces.com/gym/105386}}  
\end{itemize}

For the CCPC, we select the following four regional contests and provide their corresponding URLs:
\begin{itemize}
    \item \textbf{CCPC-HB.} CCPC Harbin Regional Contest\footnote{\url{https://codeforces.com/gym/105459}}
    \item \textbf{CCPC-JN.} CCPC Jina Regional Contest\footnote{\url{https://codeforces.com/gym/105540}}   
    \item \textbf{CCPC-ZZ.} CCPC Zhengzhou Regional Contest\footnote{\url{https://codeforces.com/gym/105632}} 
    \item \textbf{CCPC-SF.} CCPC Contest Special for Female\footnote{\url{https://codeforces.com/gym/105487}}
\end{itemize}

\subsubsection{Competitive Problem Filtering}
\label{subsubsec:competitiveproblemfiltering}

To ensure alignment with the current LLM capability, we designed the following filtering rules.

\textbf{Filter 1: Golden problem.}
In contests, the ``Golden problem" typically refers to the most challenging and complex problem. These problems often involve advanced algorithms, intricate mathematical theories, or require highly innovative problem-solving approaches~\cite{mhossain2025ll}.
Recently, LLMs have encountered significant difficulties in solving such high-difficulty problems due to their limited complex reasoning capabilities and insufficient deep mathematical understanding~\cite{huang2023competition,zheng2025livecodebench}.
Based on these considerations, we filter out this type of problem from our benchmark. For example, in ICPC-CD, problem H (Friendship is Magic)\footnote{\url{https://codeforces.com/gym/105486/problem/H}} is a golden problem, and only two competitors can pass all the test cases on Codeforces.
 
\textbf{Filter 2: Geometry problems with visual images.}
Some competitive programming problems, particularly those related to computational geometry, include diagrams or rely heavily on visual representations to convey essential geometric information. Current LLMs struggle to interpret and reason about such visual content, as they primarily process textual data and cannot extract spatial relationships from images. Based on these considerations, we filter out this type of problem from our benchmark. For example, in CCPC-JN, problem G (The Wheel of Fortune)\footnote{\url{https://codeforces.com/gym/105540/problem/G}} requires competitors to understand the accompanying image before solving the problem, which is also challenging for LLMs.

\textbf{Filter 3: The problem description contains difficult content to deal with.}
In algorithmic contests, the problem description may contain a large amount of content that is not suitable for LLM-based competitive program generation, such as excessive tables or overly long background stories. Including such content in the prompt may not only distract the model during the initial generation but also significantly reduce the effectiveness of subsequent strategies in our improvement framework, such as multi-turn dialogue or regeneration phases. Therefore, we filter out this type of problem from our benchmark. For example, in ICPC-KMR, the problem description of the problem K (Key Recovery)\footnote{\url{https://codeforces.com/gym/105588/problem/K}} is difficult to process and challenging to incorporate effectively into the prompt.

\textbf{Filter 4: Submission is not possible due to platform restrictions.}
Some problems on the Codeforces platform are only available via downloadable PDF files. For these problems, the platform does not provide a submission interface due to platform restrictions, making it impossible to submit solutions or obtain evaluation results. Therefore, we exclude such problems from our benchmark. For example, in CCPC-HB, problem B (Concave Hull)\footnote{\url{https://codeforces.com/gym/105459/problem/B}} and problem M (Weird Ceiling)\footnote{\url{https://codeforces.com/gym/105459/problem/M}} belong to this category.

\begin{table*}[htbp]
  \centering
  \scriptsize
  \caption{The number of problems filtered out by each criterion and the final number of selected problems}
    \begin{tabular}{lcccccccccc}
    \toprule
     &\textbf{ICPC-CD}& \textbf{ICPC-NJ}& \textbf{ICPC-SY}& \textbf{ICPC-KMR}& \textbf{ICPC-KMI}& \textbf{CCPC-HB}& \textbf{CCPC-JN}& \textbf{CCPC-ZZ}& \textbf{CCPC-SF} &\textbf{TOTAL}\\
    \midrule
     Initial Number&13& 13& 13& 13& 13& 13& 13& 13&13&117\\
     && & & & & & & & &\\
     Filter 1&-3& -3& -3& -1& -1& -1& -3& -4&-2&21\\
     && & & & & & & & &\\
     Filter 2&/& /& -1& /& /& /& -2& -1&-1&5\\
     && & & & & & & & &\\
     Filter 3&-1& /& -2& -2& -1& /& -1& /&-2&9\\
     && & & & & & & & &\\
     Filter 4&/& /& /& /& /& -2& /& /&/&2\\
     && & & & & & & & &\\
     Final Number&9& 10& 7& 10& 11& 10& 7& 8&8 &80\\
    \bottomrule
    \end{tabular}%
  \label{tab:filter}
\end{table*}%

We present the filtering results in Table~\ref{tab:filter}, which shows the number of problems removed by each filtering criterion and the final number of selected problems. Specifically, the initial problem pool consists of 117 problems collected from nine competitive programming contests. After applying Filter 1, we removed 21 problems; Filter 2 removed 5 problems; Filter 3 removed 9 problems; and Filter 4 removed 2 problems. Finally, 80 problems were selected to construct our benchmark.

\subsubsection{Competitive Problem Classification}

In our empirical study, we classify the selected problems from two different perspectives: problem types and problem difficulty levels.

\begin{figure}[htbp] 
    \centering 
    \includegraphics[width=0.5\textwidth]{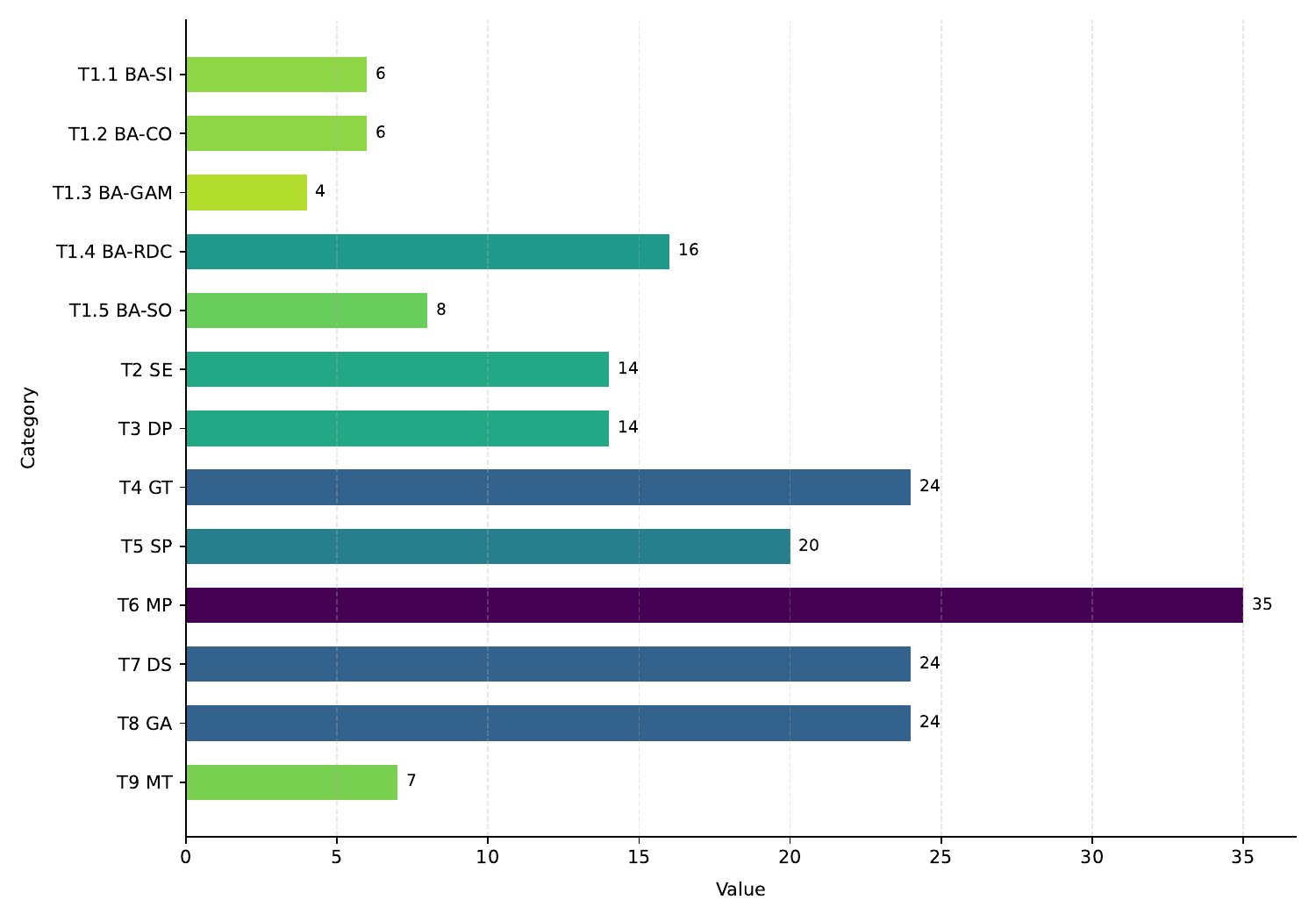}
\caption{statistics of different problem types} 
    \label{Fig:algorithmtaxonomy} 
\end{figure}

\textbf{\underline{Problem types.}} 
By following the suggestion\footnote{\url{https://oi-wiki.org/}}, we categorize the problems in our benchmark according to their problem types and report the corresponding proportions, as shown in Figure~\ref{Fig:algorithmtaxonomy}.
Note that a single problem may belong to multiple types. For example, in \textbf{ICPC-CD}, problem D (Disrupting Communications)\footnote{\url{https://codeforces.com/gym/105486/problem/E}}, the solution requires the combined use of three algorithmic techniques: graph theory, dynamic programming, and data structures.
Next, we provide a brief introduction to each problem type.

\textbf{T1 Basic Algorithms (BA).} Basic algorithms are widely applicable across various problems and are typically fundamental, concise, and conceptually straightforward.

\begin{itemize}
\item \textbf{T1.1 Simulation (SI).} Simulates specific processes or operations described in the problem.
\item \textbf{T1.2 Construction (CO).} Builds outputs directly by exploiting problem patterns or structures.
\item \textbf{T1.3 Game (GAM).} Involves decision-making under competitive or adversarial settings, often based on game theory.
\item \textbf{T1.4 Recursion \& Divide-and-Conquer (RDC).} Solves problems by recursively breaking them into smaller parts.
\item \textbf{T1.5 Sorting (SO).} Rearranges data in a specific order using classic or customized sorting methods.
\end{itemize}

\textbf{T2 Search (SE).} Explores the solution space to find optimal or valid solutions, often via exhaustive enumeration.

\textbf{T3 Dynamic Programming (DP).} Solves problems by breaking them into overlapping subproblems with optimal substructure.

\textbf{T4 Graph Theory (GT).} Involves problems modeled as graphs, focusing on node and edge relationships.

\textbf{T5 String Problems (SP).} Deals with operations or pattern matching on character sequences.

\textbf{T6 Mathematical Problems (MP).} Requires algorithmic solutions rooted in number theory, combinatorics, or probability.

\textbf{T7 Data Structure (DS).} Involves efficient data organization for fast access, updates, or queries.

\textbf{T8 Greedy Algorithm (GA).} Builds solutions step-by-step by always choosing the locally optimal option.

\textbf{T9 Miscellaneous Technique (MT).} Covers specialized or hard-to-classify techniques.

\textbf{\underline{Problem difficulties.}}  
We categorize problem difficulty into four levels: warm-up, bronze, silver, and gold. This classification is based on multiple factors, including Codeforces difficulty ratings\footnote{\url{https://codeforces.com/blog/entry/78825}}, official post-contest solutions, and the number of successful submissions. To ensure consistency and fairness, we also refer to established mappings between Codeforces ratings and USACO\footnote{\url{https://www.vplanetcoding.com/blog/usaco-codeforces}} levels, and cross-check these with the typical distribution of difficulty levels in regional contests.

Table~\ref{tab:difficulties} presents the distribution of problem difficulties in our benchmark. Specifically, the benchmark comprises 13 warm-up problems, 36 bronze-level problems, and 31 silver-level problems.

\begin{itemize}
\item \textbf{Warm-up problems}. Focus on basic programming concepts such as variables, loops, and conditions, intended to help participants quickly get familiar with the contest environment.

\item \textbf{Bronze-level problems}. Involve fundamental algorithms and data structures such as sorting and simulation. They are suitable for beginners and emphasize accurate implementation.

\item \textbf{Silver-level problems}. Require more advanced techniques such as graph traversal and prefix sums, focusing on algorithm efficiency and complexity analysis.

\item \textbf{Gold-level problems}. Involve complex algorithms like dynamic programming and computational geometry, targeting experienced participants with strong problem-solving skills. Notice in Section~\ref{subsubsec:competitiveproblemfiltering}, gold-level problems are excluded from our benchmark.
\end{itemize}

\begin{table*}[htbp]
  \centering
  \scriptsize
  \caption{Statistics of different difficulty levels across different competitive contests}
    \begin{tabular}{lcccccccccc}
    \toprule
     &\textbf{ICPC-CD}& \textbf{ICPC-NJ}& \textbf{ICPC-SY}& \textbf{ICPC-KMR}& \textbf{ICPC-KMI}& \textbf{CCPC-HB}& \textbf{CCPC-JN}& \textbf{CCPC-ZZ}& \textbf{CCPC-SF} &\textbf{TOTAL}\\
    \midrule
     Warm-up&1& 2& 1& 2& 1& 1& 2& 1&2&13\\
     && & & & & & & & &\\
     Bronze-level&4& 4& 2& 6& 5& 5& 3& 4&3&36\\
     && & & & & & & & &\\
     Silver-level&4& 3& 4& 2& 5& 5& 2& 3&3&31\\
    \bottomrule
    \end{tabular}%
  \label{tab:difficulties}
\end{table*}%

\subsection{LLM Selection}

In our empirical study, we use DeepSeek-R1 as the representative LLM to evaluate competitive program generation performance. We select DeepSeek-R1 due to its strong performance on programming benchmarks, effective instruction-following ability, and cost-efficiency for large-scale inference. According to its official documentation~\cite{guo2025deepseek}, DeepSeek-R1 achieves overall code generation performance comparable to that of ChatGPT, making it a practical choice for this task.

For DeepSeek-R1, we set the temperature hyperparameter to 0.7. This choice is inspired by recent insights into code-specific temperature adaptation presented by Zhu et al.~\cite{zhu2024hot}, which suggest that a moderate temperature can effectively balance the diversity and accuracy of generated programs.

\subsection{Basic Prompt Design}

Following the previous study~\cite{liu2024no}, we design a basic prompt tailored to the characteristics of the competitive program generation task. Specifically, our prompt consists of three components: (1) a natural language (NL) description part that includes role-playing instructions and task instructions; (2) a question stem (QS) part that provides the problem background along with input and output specifications; and (3) a test case (TC) part that presents an example input-output pair to illustrate the expected behavior. In the following, we provide a detailed description of each part.

\textbf{NL Part:} To optimize prompt design, this component incorporates two widely adopted types of instructions:
(1) role-playing instructions: By assigning the LLM a specific role (e.g., ``You are a participant in a competitive contest"), by providing a problem description above and a test case below, the model is encouraged to behave as a contestant and generate a corresponding solution.
(2) explicit task instructions: Specifying the desired outcome (e.g. ``Please use C++ programming language to solve this competitive problem") helps guide the model toward the intended task, reducing ambiguity and improving the quality of the generated code. 

\textbf{QS Part:} In alignment with prior work on LLM-based code generation~\cite{liu2024guiding}, this component includes the following elements:
(1) a comprehensive introduction to the problem background, including clarifications of any domain-specific terminology;
(2) relevant notes or constraints specified within the problem description;
(3) a detailed specification of the input format; and
(4) a detailed specification of the output format.

\textbf{TC Part:} This component contains corresponding test cases. Test cases play a crucial role in evaluating the correctness of generated programs. They not only verify whether a program meets the expected functional requirements but also help uncover potential bugs for edge cases. 

\begin{figure}[htbp] 
    \centering 
    \includegraphics[width=0.5\textwidth]{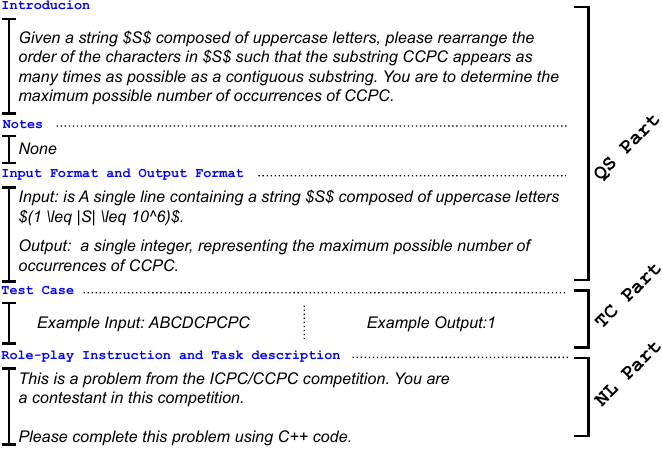}
\caption{An example to illustrate our basic prompt design} 
    \label{Fig:promptdesign} 
\end{figure}

We use an illustrative example shown in Figure~\ref{Fig:promptdesign} to demonstrate our designed basic prompt. This example corresponds to Problem C from the CCPC-SF contest\footnote{\url{https://codeforces.com/gym/105487/problem/C}}. 
Notice we utilize the ``deep thinking” feature of DeepSeek when sending the prompt to enhance the reasoning process of the model.

\begin{figure}[htbp] 
    \centering 
    \includegraphics[width=0.5\textwidth]{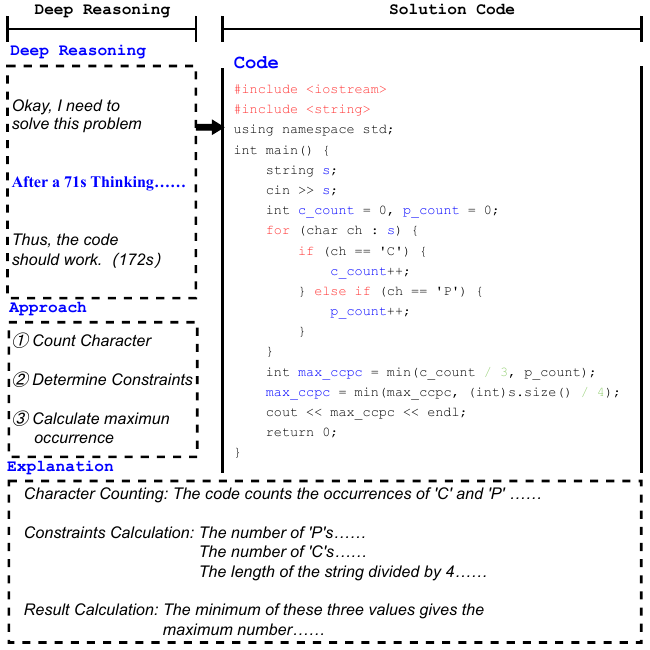}
\caption{An example to illustrate the output of LLM} 
    \label{Fig:output} 
\end{figure}

Figure~\ref{Fig:output} shows the corresponding output results of this illustrative example. The response from DeepSeek is composed of four main sections: deep thinking, approach, C++ code, and explanation. The \textbf{deep thinking} section captures the model's internal reasoning process when interpreting the prompt. The \textbf{approach} section outlines the proposed solution strategy. The \textbf{C++ code} section presents the generated algorithm implementation. Finally, the \textbf{explanation} section provides a detailed interpretation of key code segments.
 
\section{Research Questions}
\label{sec:researchquestions}

To evaluate the LLM's ability for competitive problem generation, we want to answer the following two research questions.

\textbf{RQ1: Given our designed basic prompt, how correct is the competitive program generated by LLM?}

\textbf{Motivation.} 
In RQ1, we aim to assess the basic code generation ability of LLMs on real-world competitive programming problems. Using our designed prompt, we aim to evaluate how well DeepSeek-R1 performs without additional guidance. To reveal strengths and limitations, we analyze results by problem difficulty and problem types, offering a clear foundation for understanding LLMs' capabilities on this task.

\textbf{RQ2: What types of errors commonly occur in incorrect programs generated by LLM?}

\textbf{Motivation.}  In RQ2, we aim to conduct an in-depth analysis of the incorrect solutions generated by the LLM to uncover the underlying causes of failure. By systematically examining the failed cases from RQ1, we seek to identify recurring error patterns and classify them into a comprehensive hierarchical error taxonomy. This taxonomy will help to better understand the types and distribution of errors (such as algorithm design flaws, logic mistakes, or performance bottlenecks) that limit the effectiveness of LLMs in competitive program generation. The insights gained from this analysis are crucial for designing the targeted optimization strategies for improvement.

\section{LLM Evaluation Results}
\label{sec:resultsofLLMevaluation}

\subsection{RQ1: Correctness Analysis}

\textbf{Approach.}

We utilize OJ platforms to assess the correctness of generated programs. This fully automated process enables fair, standardized, and efficient verification of both functional correctness and runtime performance.
When a generated program is submitted to an OJ platform, it may return one of the following statuses:

\begin{itemize}
    \item \textbf{AC (Accepted).} The program produces correct outputs for all test cases within the required constraints.
    \item \textbf{WA (Wrong Answer).} The program runs but generates incorrect results for one or more test cases. 
    \item \textbf{RE (Runtime Error).} The program crashes during execution, commonly due to issues such as segmentation faults or division by zero.
    \item \textbf{TLE (Time Limit Exceeded).} The program fails to complete execution within the prescribed time limit.
    \item \textbf{CE (Compile Error).} The program fails to compile due to syntax errors or unsupported language feature usage.
\end{itemize}

\textbf{Results.}
To evaluate the effectiveness of DeepSeek-R1 under our designed basic prompt, we submitted the generated solutions for all 80 competitive programming problems in our benchmark. The results show that only 5 solutions achieved AC status, successfully passing all test cases. In contrast, 63 submissions resulted in WA, 5 in TLE, and 7 in CE, indicating that most of the generated programs failed to produce correct or efficient outputs.
This distribution highlights the substantial limitations of current LLMs in handling real-world algorithmic challenges when only using the basic prompt. To gain deeper insights, we further analyze the detailed results from two different perspectives: problem difficulty levels and problem types.

\textbf{\underline{Analysis in terms of difficulty levels.}}

The results by difficulty level are shown in Table~\ref{tab:beforecomparison1}. Specifically, among the 13 warm-up problems, 4 submissions achieved AC, 1 resulted in TLE, and the remaining 8 were WA. For the 36 bronze-level problems, there were 0 AC, 1 TLE, 3 CE, and 32 WA. Among the 31 Silver-level problems, only 1 achieved AC, while 3 resulted in TLE, 4 in CE, and 23 in WA. These results suggest that DeepSeek-R1 performs better on low-difficulty problems, while its effectiveness drops sharply on more challenging ones, highlighting its limitations in handling high-difficulty problems.

\begin{table}[htbp]
  \centering
  \scriptsize
  \caption{Program generation result analysis in terms of problem difficulty level with basic prompt}

    \begin{tabular}{>{\centering\arraybackslash}p{2cm}>{\centering\arraybackslash}p{1cm}>{\centering\arraybackslash}p{4.5cm}}
    \toprule
    \textbf{Difficulty Level} & \textbf{AC} & \textbf{Other Status} \\
    \midrule
    Warm-up & 4/13& 9/13 (8 WA, 1 TLE)\\
    &&\\
    Bronze & 0/36& 36/36 (32 WA, 1 TLE, 3 CE)\\
    &&\\
    Silver & 1/31& 30/31 (23 WA, 3 TLE, 4 CE)\\
    \midrule
    \textbf{Summary} & 5/80& 75/80 (63 WA, 5 TLE, 7 CE)\\
    \bottomrule
    \end{tabular}
  \label{tab:beforecomparison1}
\end{table}

\textbf{\underline{Analysis in terms of problem types.}}

Notice to facilitate analysis, we briefly categorize the problems into single-algorithm and multi-algorithm types based on the number of algorithms involved.
The results are presented in Table~\ref{tab:beforecomparison2}. 
Among the 49 single-algorithm problems, 4 submissions achieved AC, 37 resulted in WA, 4 in TLE, and 37 were CE. In contrast, for the 31 multi-algorithm problems, only 1 submission achieved AC, with 26 WA, 1 TLE, and 3 CE. 
These results indicate that LLMs, such as DeepSeek-R1, perform significantly better on single-algorithm problems compared to multi-algorithm ones, suggesting that solving problems requiring the integration of multiple algorithms poses a greater challenge for competitive code generation.

\begin{table}[htbp]
  \centering
  \scriptsize
  \caption{Program generation result analysis in terms of problem types (i.e., single-algorithm type and multi-algorithm type) with basic prompt}
  \begin{tabular}{>{\centering\arraybackslash}p{2cm}>{\centering\arraybackslash}p{1cm}>{\centering\arraybackslash}p{4.5cm}}
    \toprule
    \textbf{Problem Type} & \textbf{AC} & \textbf{Other Status} \\
    \midrule
    Single algorithm& 4/49& 45/49 (37 WA, 4 TLE, 4 CE)\\
    &&\\
    Multi-algorithm& 1/31& 30/31 (26 WA, 1 TLE, 3 CE)\\
    \midrule
    \textbf{Summary} & 5/80& 75/80 (63 WA, 5 TLE, 7 CE)\\
    \bottomrule
  \end{tabular}
  \label{tab:beforecomparison2}
\end{table}%

\begin{tcolorbox}[width=1.0\linewidth, title={Summary for RQ1:}]

LLMs demonstrate limited effectiveness on competitive programming tasks when using basic prompts, with only 5 out of 80 submissions achieving AC.

Submission correctness declines with increasing difficulty, and multi-algorithm problems result in more errors, indicating LLMs struggle with both algorithmic complexity and integration of multiple algorithms.

\end{tcolorbox}

\subsection{RQ2: Error Taxonomy Construction}

\textbf{Approach.} 
To construct a fine-grained taxonomy of errors, we randomly selected 80\% of the non-AC (not accepted) programs (i.e., 60 out of 75 incorrect submissions) as our development set. This taxonomy was constructed by following the previous study~\cite{wen2021empirical,chen2025empirical}. The annotation was conducted collaboratively by two annotators (i.e., the fifth and sixth authors), both of whom are top performers from a university algorithm training team. They carefully examined each program's structure, including function calls, entry points, output behavior, and algorithmic logic, to perform error diagnosis on competitive algorithmic programs generated by LLMs.

We adopted an open coding procedure~\cite{seaman1999qualitative}, wherein each annotator reviewed all development set programs at least twice. During this analysis, they identified and labeled all observable errors, summarizing each one using concise and descriptive phrases. Some programs exhibited multiple distinct errors contributing to test failures, for example, a single solution might simultaneously suffer from ``misunderstanding problem requirements'' and ``failure to handle boundary conditions.'' For these cases, the program was first debugged and manually corrected into an AC version; suspected faulty regions were then re-inserted and validated incrementally to pinpoint all the errors leading to failure.

The annotators iteratively refined the taxonomy by processing each erroneous program and updating error descriptions to improve clarity and granularity. If a program contained multiple types of errors, each was independently assigned to the corresponding category. In cases of disagreement, a third adjudicator (i.e., the third author), who serves as the leader of the algorithm training team, resolved conflicts through discussion. After thorough inspection and agreement among all participants, we finalized a preliminary error taxonomy covering the full spectrum of LLM-generated failure modes.

Subsequently, we applied the preliminary taxonomy to the remaining 20\% of incorrect programs as a validation set. The same two annotators independently labeled these 15 programs by inspecting the code, identifying faults via backfilling, and assigning each error to its corresponding category. If an error could not be classified, it was temporarily labeled as ``Pending.'' We used Cohen’s Kappa ($\kappa$)~\cite{cohen1960coefficient} to evaluate inter-annotator agreement, yielding a score of 0.86, which indicates excellent reliability~\cite{landis1977measurement} and confirms the robustness of our open coding procedure.

For ``Pending'' cases, we brought in a fourth annotator (i.e., the fourth author) to mediate and assist in determining the appropriate categories. If a previously unobserved error type emerged during this step, we added new categories accordingly. In total, six algorithm-specific error categories were introduced to extend the taxonomy. The final taxonomy thus comprehensively captures the diverse error patterns present in LLM-generated competitive programs, with consensus reached among all annotators. Detailed results are reported in Section~\ref{sec:rq2result}.

    \begin{figure*}[htbp] 
        \centering 
        \includegraphics[width=0.90\textwidth]{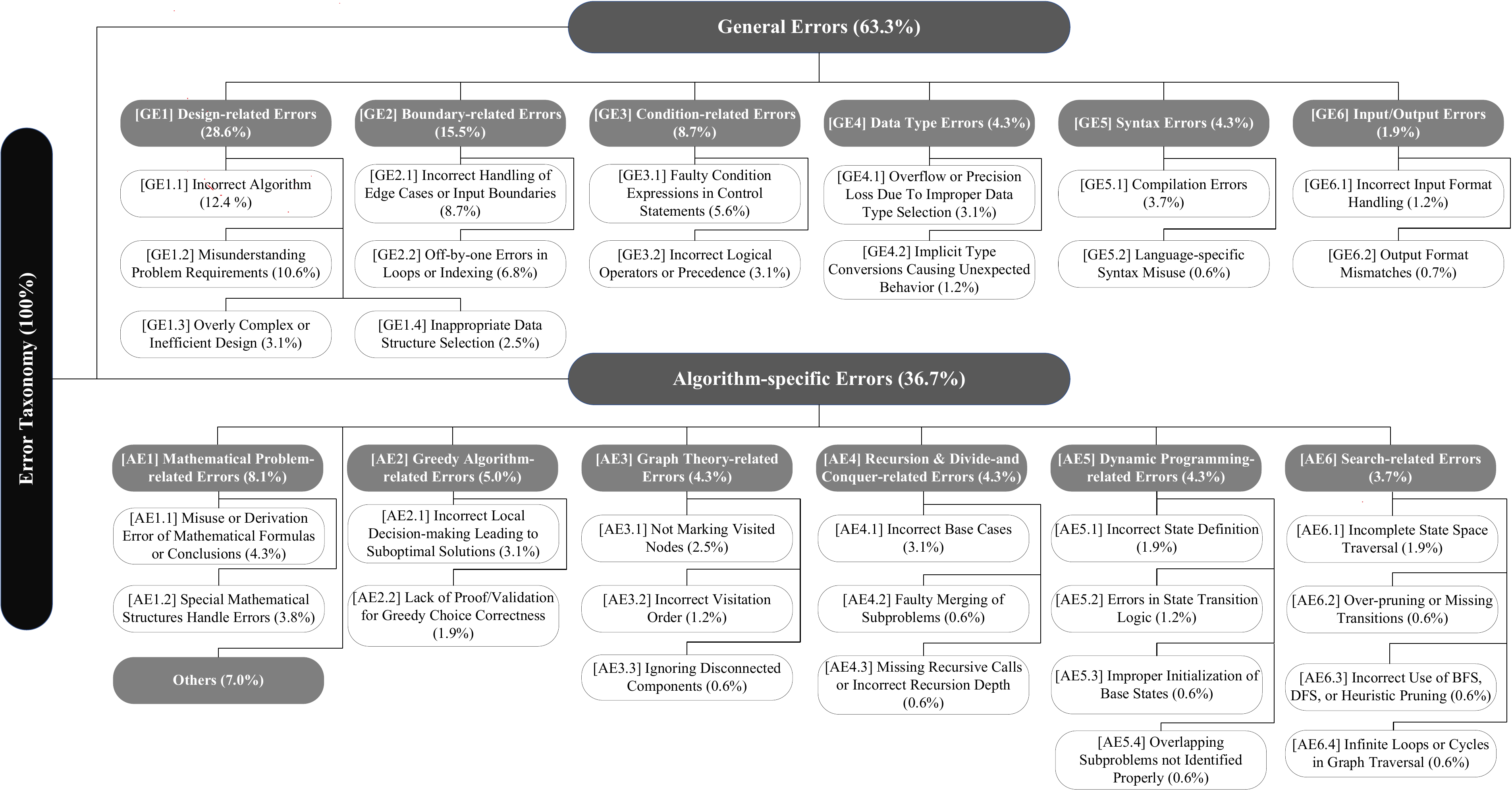}
    \caption{Hierarchical error taxonomy of LLM-generated competitive programs.} 
        \label{Fig:ErrorTaxonomy} 
    \end{figure*}

\textbf{Result.}
\label{sec:rq2result}

\begin{figure}[htbp]
     \centering 
     \includegraphics[width=0.45\textwidth]{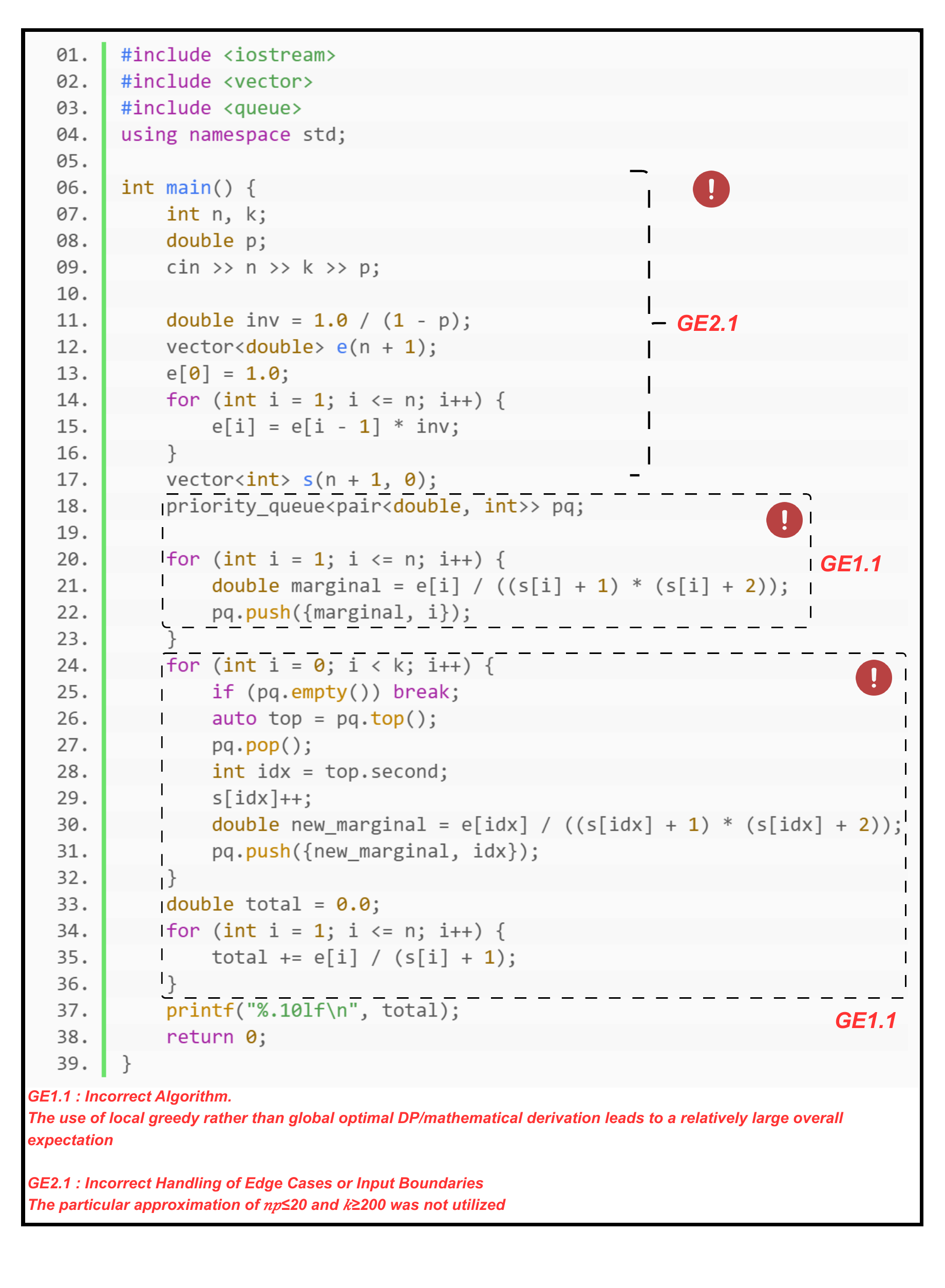}
     \caption{An example of the generated solution involving multiple error categories.} 
     \label{Fig:scpimec} 
     \end{figure}

We use Figure~\ref{Fig:ErrorTaxonomy} to present our constructed hierarchical error taxonomy. Specifically, it consists of two main categories: general errors and algorithm-specific errors. For each category, we report its corresponding proportion. Note that a single erroneous program may involve multiple types of errors, and thus can be associated with multiple categories. 
To better illustrate this, we provide a representative example (Problem K~\footnote{\url{https://codeforces.com/gym/105459/problem/C}} from CCPC-ZZ) in Figure~\ref{Fig:scpimec}.
Then we provide detailed explanations for each category.

\textbf{\underline{General Errors.}}
General errors capture various aspects of program correctness, encompassing design-related errors, syntax errors, input/output errors, boundary-related errors, condition-related errors, and data type errors. Then we provide detailed descriptions of each top-level category along with its corresponding subcategories.

\textbf{Design-related Errors (GE1).} 
These errors stem from flaws in the overall solution strategy or algorithm design, often caused by a misunderstanding of the problem requirements or the selection of an inappropriate algorithm. Fixing such errors typically requires substantial revisions to the algorithm design, rather than simple code-level modifications.
This category consists of four subcategories.

\begin{itemize}
   \item \textbf{Incorrect Algorithm (GE1.1).}  
This type of error arises when the model selects an inappropriate algorithm for the given problem, leading to a fundamental flaw in the overall solution strategy.
A common example is the use of a greedy algorithm for problems that inherently require dynamic programming or exhaustive search. While a greedy approach might pass simple test cases, it often fails on edge cases due to its inability to account for all of the problem’s constraints.

    \item \textbf{Misunderstanding Problem Requirements (GE1.2).}
This type of error stems from the model’s incorrect semantic understanding of the problem description. Such misunderstandings typically occur during the initial problem analysis phase, which leads to the subsequent algorithm design in the wrong direction. As a result, even if the implementation is logically consistent, the final program fails to fulfill the actual requirements of the task.
An illustrative example can be found in Figure~\ref{Fig:KMI-G}, which features Problem G from ICPC-KMI~\footnote{\url{https://codeforces.com/gym/105386/problem/G}}. In the figure, the upper section displays the problem description, while the lower section presents an incorrect code snippet accompanied by a brief analysis of the corresponding error on the side.

     \begin{figure}[htbp]
     \centering 
     \includegraphics[width=0.5\textwidth]{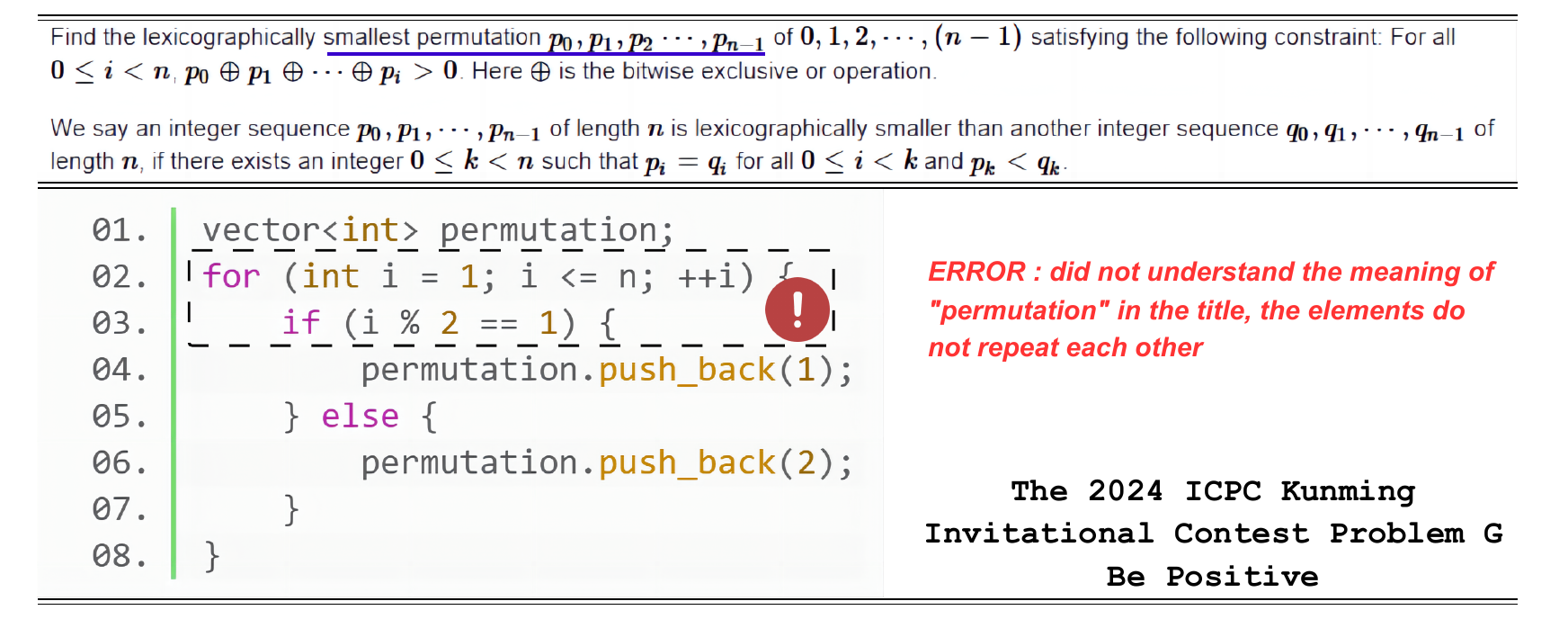}
     \caption{An example of GE1.2 error} 
     \label{Fig:KMI-G} 
     \end{figure}

    \item \textbf{Overly Complex or Inefficient Design (GE1.3).}  
This type of error refers to situations where the solver constructs a solution that, while logically correct, incorporates unnecessary layers of computation, overly complex control flow, or inefficient algorithms. Such designs tend to increase time or space complexity, making the solution susceptible to time-limit or memory-limit violations, especially when handling large-scale inputs.
An illustrative example can be found in Figure~\ref{Fig:KMI-M}, which showcases Problem M from ICPC-KMI~\footnote{\url{https://codeforces.com/gym/105386/problem/M}}.

     \begin{figure}[htbp]
     \centering 
     \includegraphics[width=0.5\textwidth]{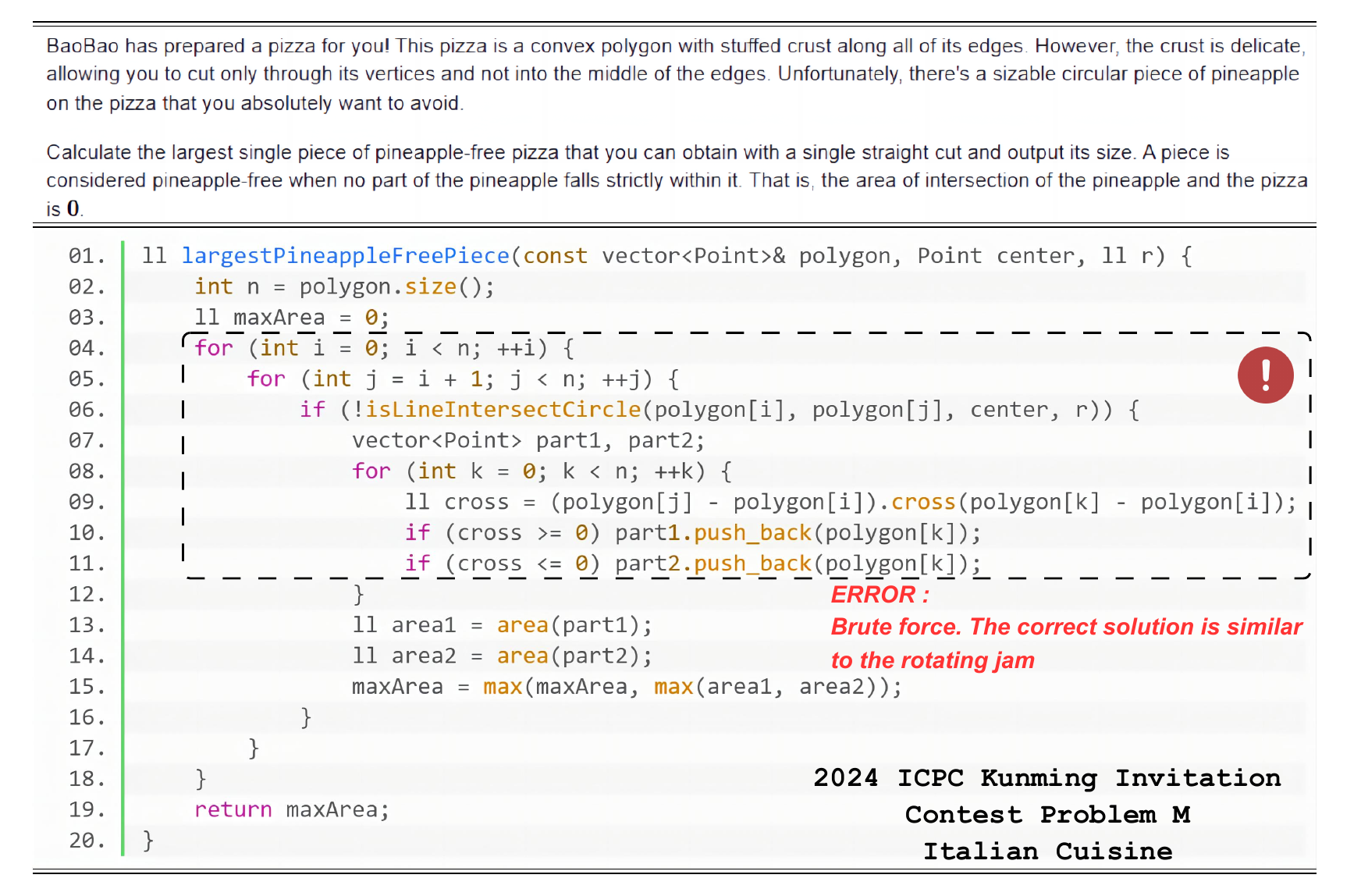}
     \caption{An example of GE1.3 error} 
     \label{Fig:KMI-M} 
     \end{figure}

    \item \textbf{Inappropriate Data Structure Selection (GE1.4).}  
This type of error arises when the selected data structure does not align with the operational requirements or constraints of the problem. Such a mismatch can lead to low time or space efficiency and may even prevent the correct implementation of the intended algorithm, particularly in problems with strict performance constraints.

\end{itemize}

\textbf{Boundary-related Errors (GE2).} Improper handling of edge values can compromise the program’s logical correctness at the boundaries of the input domain. These errors often remain undetected until exposed by adversarial or hidden test cases.

\begin{itemize}
    \item \textbf{Incorrect Handling of Edge Cases or Input Boundaries (GE2.1).}
This type of error occurs when the program fails to properly handle edge cases or input boundaries, leading to incorrect behavior under extreme or uncommon conditions
An example (Problem A~\footnote{\url{https://codeforces.com/gym/105540/problem/A}} from CCPC-JN) can be found in Figure~\ref{Fig:JN-A}.

    \begin{figure}[htbp]
     \centering 
     \includegraphics[width=0.5\textwidth]{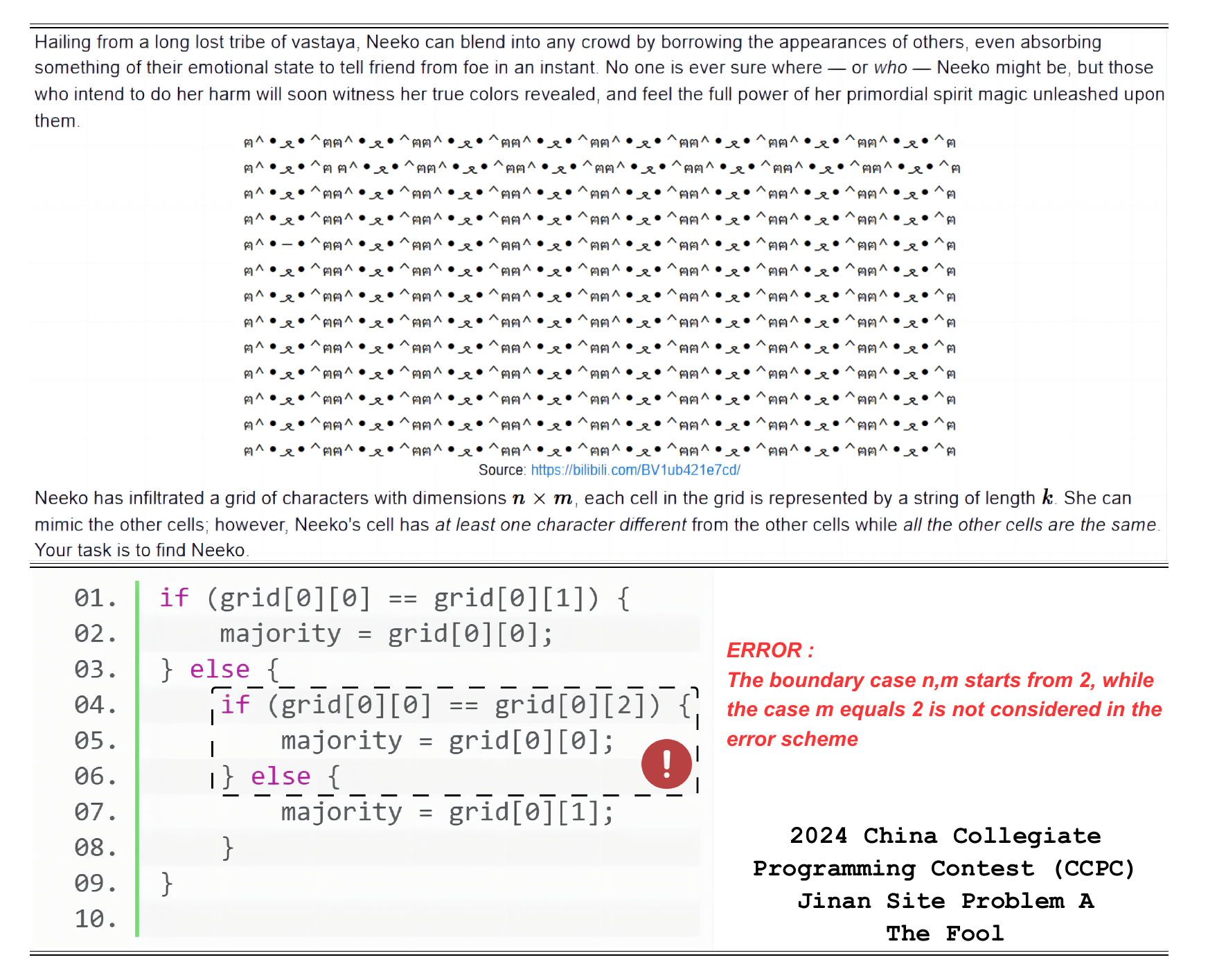}
     \caption{An example of GE2.1 error} 
     \label{Fig:JN-A} 
     \end{figure}

    \item \textbf{Off-by-one Errors in Loops or Indexing (GE2.2).}
This type of error arises when loop boundaries or array indices are set incorrectly, typically due to using \texttt{<} instead of \texttt{<=}, or starting from an off-position such as 1 instead of 0. These errors often lead to out-of-bounds access, skipped elements, or incomplete processing of edge cases, especially in problems involving arrays, strings, or intervals.
An example (Problem A~\footnote{\url{https://codeforces.com/gym/105386/problem/A}} from ICPC-KMI) can be found in Figure~\ref{Fig:KMI-A}.

     \begin{figure}[htbp]
     \centering 
     \includegraphics[width=0.5\textwidth]{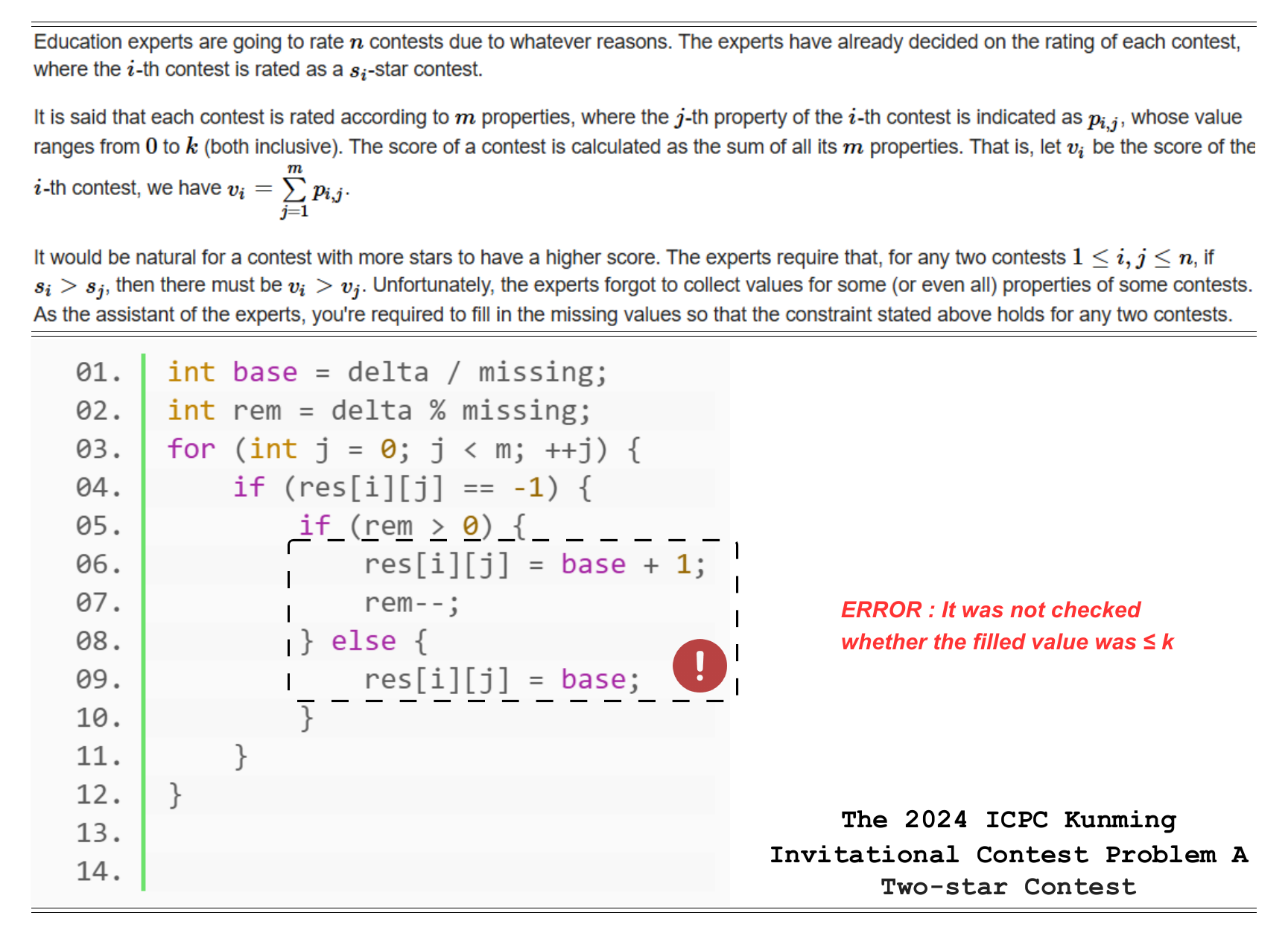}
     \caption{An example of GE2.2 error} 
     \label{Fig:KMI-A} 
     \end{figure}

\end{itemize}

\textbf{Condition-related Errors (GE3).} Faulty boolean expressions (such as incorrect operators, misinterpreted precedence, or misplaced negations) can misdirect control flow or prematurely terminate loops. These errors typically result in partially correct outputs or missed corner cases.

\begin{itemize}
    \item \textbf{Faulty Condition Expressions in Control Statements (GE3.1).}
This type of error occurs when the conditions within control structures are incorrectly formulated, causing essential execution paths to be skipped or irrelevant code blocks to be executed unintentionally.
An example (Problem C~\footnote{\url{https://codeforces.com/gym/105459/problem/C}} from CCPC-HB) can be found in Figure~\ref{Fig:HB-C}.

    \begin{figure}[htbp]
     \centering 
     \includegraphics[width=0.45\textwidth]{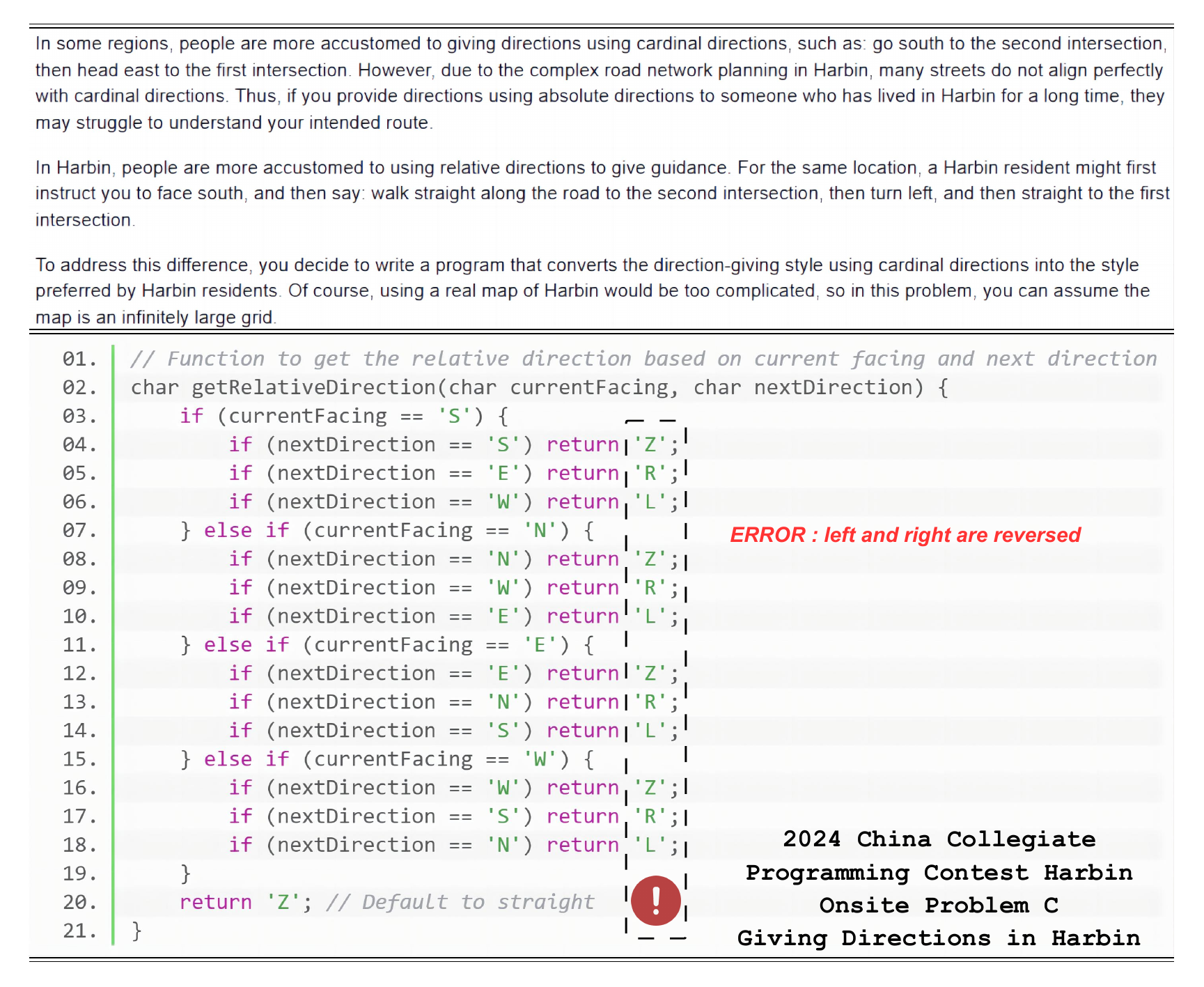}
     \caption{An example of GE3.1 error}  
     \label{Fig:HB-C} 
     \end{figure}

    \item \textbf{Incorrect Logical Operators or Precedence (GE3.2).}
This type of error arises when logical expressions use incorrect operators or depend on operator precedence without proper use of parentheses. As a result, control flow may follow unintended branches, leading to missed conditions or incorrect outputs in test cases.
\end{itemize}

\textbf{Data Type Errors (GE4).} 
This error occurs when an inappropriate numeric range or precision is used, or when implicit type conversions alter the intended semantics. Such issues can lead to overflow, underflow, or subtle rounding errors. Selecting an appropriate data type or applying explicit casts can effectively resolve the problem.

\begin{itemize}
    \item \textbf{Overflow or Precision Loss Due To Improper Data Type Selection (GE4.1).}
This error arises when a variable is declared with a data type that lacks sufficient range or precision to represent the required values. For instance, using \texttt{int} where \texttt{long long} is necessary can cause integer overflow and lead to incorrect results.

    \item \textbf{Implicit Type Conversions Causing Unexpected Behavior (GE4.2).}
This error occurs when operations involve mixed data types, causing implicit conversions that alter the program’s intended semantics.  Such conversions can lead to loss of precision, sign mismatches, or unintended value truncation, particularly in arithmetic or conditional expressions.
\end{itemize}

\textbf{Syntax Errors (GE5).} Purely lexical or grammatical mistakes that prevent compilation or produce immediate runtime faults, such as missing tokens, mismatched delimiters, or language-specific misuse of keywords. They reflect lapses in basic code authoring rather than conceptual misunderstanding.

\begin{itemize}
    \item \textbf{Compilation Errors (GE5.1).}
Syntax errors occur when the program fails to compile due to missing semicolons, mismatched brackets, undeclared variables, or other programming language syntax rule violations.

    \item \textbf{Language-specific Syntax Misuse (GE5.2).}
This refers to the incorrect use of programming language features that behave differently depending on the language environment. For example, in C++, mixing \texttt{cin}/\texttt{cout} with \texttt{scanf}/\texttt{printf} without disabling synchronization (via \texttt{std::ios::sync\_with\_stdio(false)} and \texttt{cin.tie(nullptr)}) can lead to unexpected I/O behavior or significant performance degradation. Such misuse does not cause compilation errors but can result in inefficient execution.
\end{itemize}

\textbf{Input/Output Errors (GE6).} The program mishandles the prescribed I/O contract: it misparses input, formats output incorrectly, or neglects special cases like empty streams, leading to WA status despite correct internal logic. 

\begin{itemize}
    \item \textbf{Incorrect Input Format Handling (GE6.1).}
This error occurs when the program does not adhere to the input format specified in the problem statement. As a result, it may read input incorrectly or prematurely, leading to incorrect values being processed by the subsequent logic.

    \item \textbf{Output Format Mismatches (GE6.2).}
This type of error occurs when the program's output format does not strictly conform to the problem's specifications. Even if the underlying logic and computed values are correct, such formatting issues can still lead to a WA judgment.
An example (Problem D~\footnote{\url{https://codeforces.com/gym/105578/problem/D}} from ICPC-SY) can be found in Figure~\ref{Fig:SY-D}.

  \begin{figure}[htbp]
     \centering 
     \includegraphics[width=0.5\textwidth]{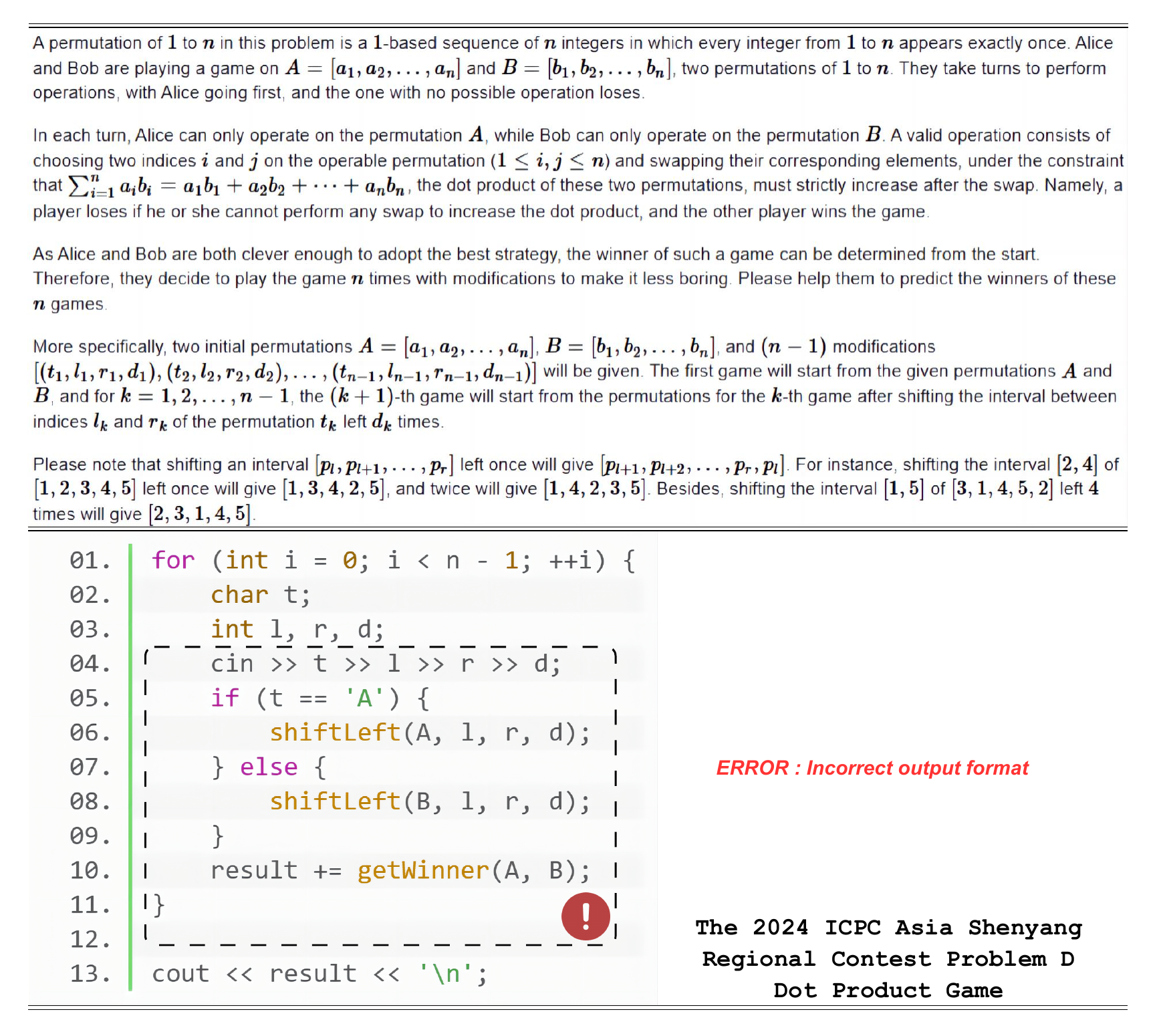}
     \caption{An example of GE6.2 error} 
     \label{Fig:SY-D} 
     \end{figure}
\end{itemize}

\textbf{\underline{Algorithm-specific Errors.}} 
Algorithm-specific errors refer to logic-specific mistakes tied to particular algorithm types (e.g., dynamic programming, graph algorithms, greedy algorithms)

\textbf{Mathematical Problem-related Errors (AE1).}  Errors related to mathematical problems.

\begin{itemize}
    \item \textbf{Misuse or Derivation Error of Mathematical Formulas or Conclusions (AE1.1).}  
These errors occur when the solver applies mathematical formulas incorrectly, derives flawed expressions, or ignores the necessary preconditions for applying known results. Common issues include misusing modular arithmetic identities, applying combinatorial formulas without handling edge conditions, or invoking theorems under invalid assumptions.

    \item \textbf{Special Mathematical Structure Handle Errors (AE1.2).}  
When dealing with data with special mathematical properties (such as prime sieve, factorization, GCD, modular inverses), the model often makes mistakes in implementation, has logical structure defects, or fails to cover all special cases.
\end{itemize}

\textbf{Greedy Algorithm-related Errors (AE2).} Errors related to the greedy algorithm.

\begin{itemize}
    \item \textbf{Incorrect Local Decision-making Leading to Suboptimal Solutions (AE2.1).}
Greedy algorithms make choices based on local optimality with the hope that this leads to a global optimum. When the local decision rule is flawed, such as picking the largest item without considering constraints, it may yield a feasible but non-optimal solution.

    \item \textbf{Lack of Proof/Validation for Greedy Choice Correctness (AE2.2).}
Greedy algorithms require a correctness guarantee (e.g., the property of greedy choice or the optimal substructure). Failing to validate that the greedy strategy works for all inputs formally can lead to solutions that pass sample cases but fail on edge cases, especially where greedy behavior breaks down.
\end{itemize}

\textbf{Graph Theory-related Errors (AE3).} Errors related to graph-theory algorithms.

\begin{itemize}
\item \textbf{Not Marking Visited Nodes (AE3.1).}
Failing to mark nodes as visited during traversal may lead to infinite loops or repeated visits, particularly in graphs with cycles. This often breaks DFS or BFS logic and affects termination. 

\item \textbf{Incorrect Visitation Order (AE3.2).}  
Marking nodes too late or processing them in the wrong sequence can cause incorrect behavior in traversal-based algorithms (i.e., topological sort, cycle detection). Ensuring proper visitation timing is key to correctness. 

\item \textbf{Ignoring Disconnected Components (AE3.3).}  
Only exploring from a single start node and neglecting other unvisited nodes can cause entire components to be missed. This results in incomplete outputs for problems involving connectivity or component counting.

\end{itemize}

\textbf{Recursion \& Divide-and-Conquer-related Errors (AE4).} Errors related to recursion and divide-and-conquer.

\begin{itemize}
    \item \textbf{Incorrect Base Cases (AE4.1).}
In divide-and-conquer algorithms, recursion must terminate at well-defined base cases. Errors in base case logic, such as incorrect stopping conditions or failure to handle minimal input sizes, can result in infinite recursion or wrong answers on trivial subproblems.

\item \textbf{Faulty Merging of Subproblems (AE4.2).}
After dividing the problem and solving each part recursively, the merge step is responsible for combining the sub-results into a complete solution. If the merging logic is incorrect, the final output may fail to reflect the correct solution to the original problem accurately.

    \item \textbf{Missing Recursive Calls or Incorrect Recursion Depth (AE4.3).}
Divide-and-conquer relies on recursive decomposition of the input.  Omitting necessary recursive calls or failing to reach sufficient depth can cause parts of the problem to remain unsolved, leading to incomplete or incorrect results.

\end{itemize}

\textbf{Dynamic Programming-related Errors (AE5).} Errors related to the dynamic programming.

\begin{itemize}
    \item \textbf{Incorrect State Definition  (AE5.1).}
Dynamic programming relies on defining problem states that capture sufficient information for recurrence. If the state is too coarse (i.e., loses key distinctions) or too fine (i.e., leads to over-complexity), the DP formulation will fail to represent the full problem correctly, resulting in incorrect or incomplete solutions.

    \item \textbf{Errors in State Transition Logic (AE5.2).}
State transitions describe how a larger problem is built from smaller subproblems. Mistakes in this recurrence, such as using the wrong indices, conditions, or transition direction, cause the DP table to be filled incorrectly, producing wrong final answers even when the states are defined properly.

    \item \textbf{Improper Initialization of Base States (AE5.3).}
Dynamic programming solutions depend on correctly initializing base cases (e.g., dp[0], dp[1]) from which all other values are derived. If base values are missing or set incorrectly, subsequent transitions accumulate errors, propagating incorrect values throughout the DP table.

   \item \textbf{Overlapping Subproblems not Identified Properly (AE5.4).}
A key idea of Dynamic programming is recognizing and reusing solutions to overlapping subproblems. If the problem is solved repeatedly for the same input (e.g., in a naive recursive form), it indicates that the overlapping substructure has not been exploited. This often leads to redundant computation and inefficiency.

\end{itemize}

\textbf{Search-related Errors (AE6).} Errors related to the search algorithm.

\begin{itemize}
    \item \textbf{Incomplete State Space Traversal (AE6.1).}
This error occurs when the search algorithm fails to explore all reachable states due to early termination or limited expansion logic. It leads to missing valid solutions, especially in problems requiring full coverage or global optimality.
An example (Problem B\footnote{\url{https://codeforces.com/gym/105632/problem/B}} from CCPC-ZZ) can be found in Figure~\ref{Fig:ZZ-B}.

    \begin{figure}[htbp]
     \centering 
     \includegraphics[width=0.5\textwidth]{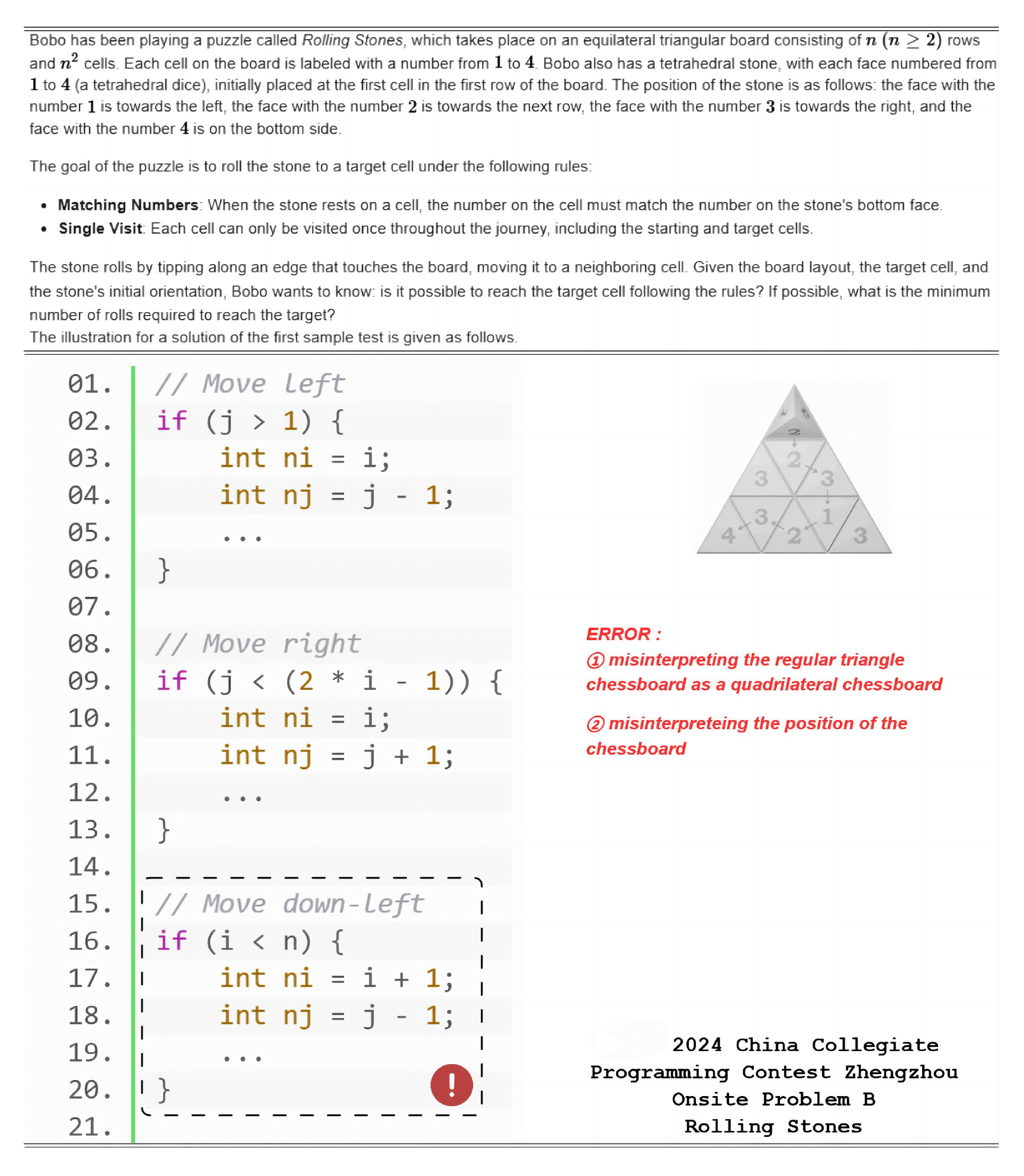}
     \caption{An example of AE4.1 error} 
     \label{Fig:ZZ-B} 
     \end{figure}

    \item \textbf{Over-pruning or Missing Transitions (AE6.2).}
Search algorithms often include pruning to improve efficiency, but overly aggressive or incorrect pruning can remove valid paths. Similarly, neglecting certain transitions in the state graph results in an incomplete or incorrect traversal.

    \item \textbf{Incorrect Use of BFS, DFS, or Heuristic Pruning (AE6.3).}
Each search strategy has a specific application scenario. 
For example, BFS guarantees the shortest path in unweighted graphs, while DFS does not. Applying the wrong strategy or misusing heuristics (e.g., in A*) can produce logically incorrect or suboptimal results.
    
    \item \textbf{Infinite Loops or Cycles in Graph Traversal (AE6.4).}
Failing to track visited states or cycles can cause the algorithm to revisit the same nodes indefinitely. This often results in non-terminating execution or redundant computation, especially in graphs with cycles or bidirectional edges.
\end{itemize}

\begin{tcolorbox}[width=1.0\linewidth, title={Summary for RQ2:}]

The error taxonomy system consists of two main categories: general errors (63.3\%) and algorithm-specific errors (36.7\%).   Within general errors, the most prevalent types are design-related errors (28. 6\%), including incorrect algorithm selection (12. 4\%) and misunderstanding problem requirements (10. 6\%), followed by boundary-related errors (15. 5\%) and condition-related errors (8. 7\%).   These errors typically arise from flawed logic, misunderstanding of the problem constraints, or inadequate control flow structures.

In contrast, algorithm-specific errors are distributed across typical algorithmic paradigms. In particular, mathematical problem-related errors (8.1\%) are the most frequent in this category, followed by greedy algorithm errors (5.0\%).   Other algorithm-specific issues, such as those involving graph theory (4.3\%), recursion \& divide-and-conquer (4.3\%), dynamic programming (4.3\%), and search algorithms (3.7\%), each contribute modest portions.
 
\end{tcolorbox}

\section{Improvement Framework}
\label{sec:LLMimprovement}

\begin{figure*}[htbp] 
    \centering 
    \includegraphics[width=0.95\textwidth]{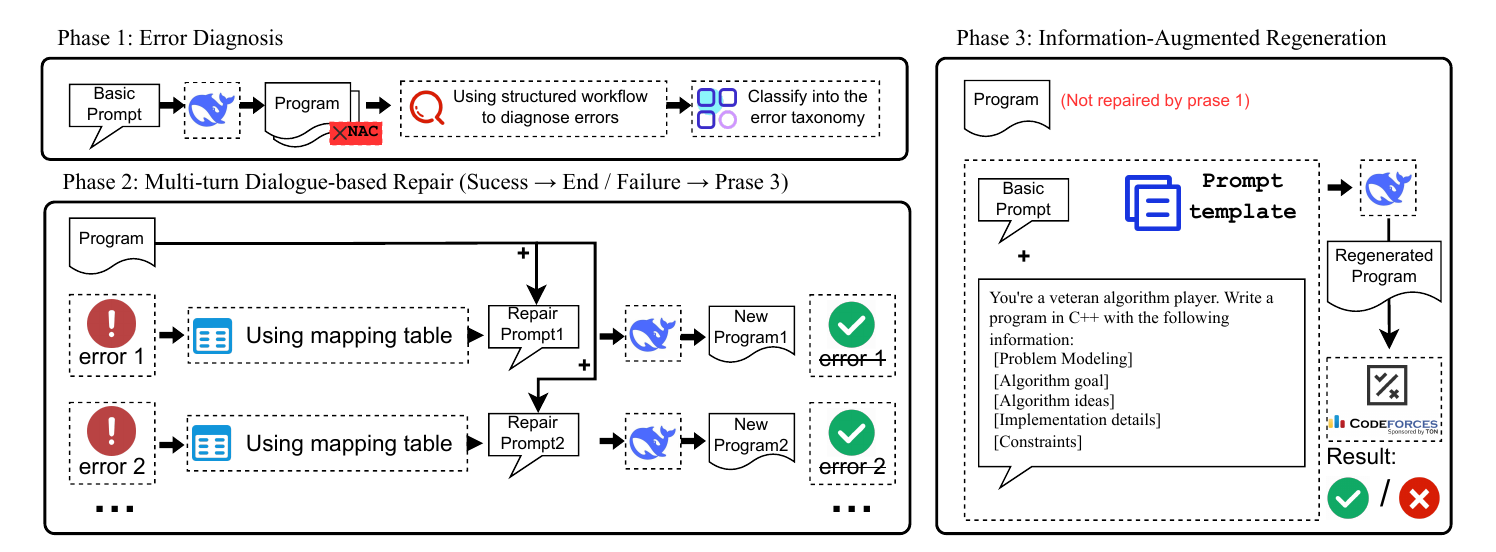}
\caption{Improvement framework for LLM-based competitive program generation} 
    \label{Fig:improvingframework} 
\end{figure*}

The evaluation results in Section~\ref{sec:resultsofLLMevaluation} reveal that most LLM-generated programs fail due to algorithm design flaws, logic errors, and improper handling of edge cases. These common failure patterns highlight the need for a more systematic approach to enhance model correctness. Motivated by these findings, we simulate the reasoning process of contestants and propose a taxonomy-driven improvement framework for LLM-generated competitive programs. The overall framework is shown in Figure~\ref{Fig:improvingframework}. consisting of three main phases:
Specifically, \textbf{Phase 1} identifies the root causes of failure and classifies errors based on a hierarchical taxonomy. \textbf{Phase 2} uses multi-turn dialogue with targeted prompts to iteratively fix the code. If the program is still incorrect, \textbf{Phase 3} regenerates a new solution using a structured prompt that incorporates key problem insights. Together, these phases enable systematic and effective improvement of LLM-generated competitive programs.

\subsection{Phase 1: Error Diagnosis}  
This phase aims to diagnose why a basic prompt-generated program fails to achieve AC and to map the observed errors into a structured taxonomy for guiding subsequent repair. It serves as the foundation of the improvement framework by ensuring that each failure is clearly understood before attempting correction.

\begin{figure*}[htbp] 
    \centering 
    \includegraphics[width=0.95\textwidth]{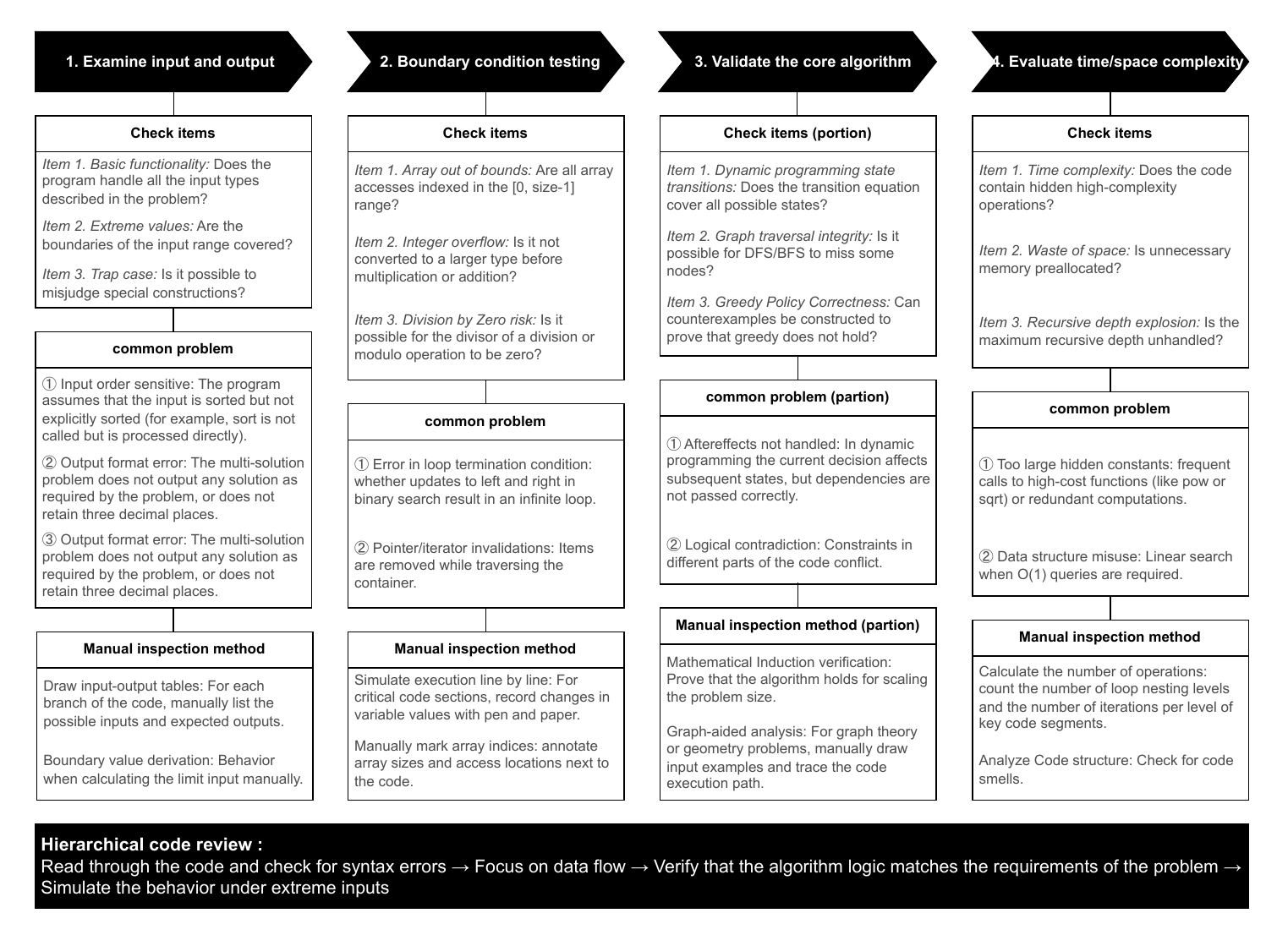}
\caption{A structured workflow for identifying and diagnosing errors in non-AC programs.} 
    \label{Fig:faxiancuowu} 
\end{figure*}

We first identify the root cause of failure using a structured inspection process, as illustrated in Figure~\ref{Fig:faxiancuowu}. This process integrates both manual and automated analysis to locate faults at various levels of the program, from high-level algorithm design to low-level implementation details. Key steps include checking functional behavior, boundary conditions, algorithm correctness, and resource efficiency. These checks help uncover common issues such as logical contradictions, format violations, memory errors, and time complexity bottlenecks.

After locating the errors, they are categorized using our fine-grained hierarchical taxonomy, which distinguishes between GE and AE. This classification not only standardizes the understanding of error types but also enables precise mapping to targeted repair strategies in the repair phases.

\subsection{Phase 2: Multi-turn Dialogue-based Repair.} 

This phase aims to iteratively repair non-AC programs by guiding the LLM through a sequence of error-aware prompts. Based on the identified error types, the framework dynamically selects appropriate repair strategies from a modular set of seven options.

In this phase, the LLM is prompted through multiple concise and targeted instructions to correct errors step-by-step. Each repair strategy is designed to address a specific class of failure. The detailed prompt templates and the mapping between error types and repair strategies are provided on GitHub~\footnote{\url{https://github.com/minnanWei/LLMs-Competitive-Program-Generation/blob/main/README.md?plain=1}}.

\begin{itemize}
    \item \textit{\textbf{Strategy 1: Full Algorithm Regeneration}}  
    Reconstruct the solution from scratch when the original algorithm paradigm is fundamentally flawed or misaligned with the problem requirements.

    \item \textbf{\textit{Strategy 2: Logic Completion}}  
    Fill in missing steps or incomplete branches in control logic, especially in conditionals or loops.

    \item \textbf{\textit{Strategy 3: Prompt-Level Clarification}}  
    Clarify ambiguous problem descriptions or constraints at the prompt level to better align model understanding with problem intent.

    \item \textbf{\textit{Strategy 4: I/O Format Fix}}  
    Standardize input/output handling to conform with expected formats (e.g., newline characters, spacing, or value delimiters).

    \item \textbf{\textit{Strategy 5: Syntax-Level Repair}}  
    Resolve compilation or syntax issues such as missing semicolons, incorrect function declarations, or unmatched brackets.

    \item \textbf{\textit{Strategy 6: Calculation Precision / Memory Safety Enhancement}}  
    Address issues related to integer overflow, floating-point errors, or unsafe memory access patterns.

    \item \textbf{\textit{Strategy 7: Termination Assurance}}  
    Modify loop boundaries or recursive calls to ensure the program halts under all input conditions.
\end{itemize}

 \begin{figure}[htbp] 
        \centering 
        \includegraphics[width=0.5\textwidth]{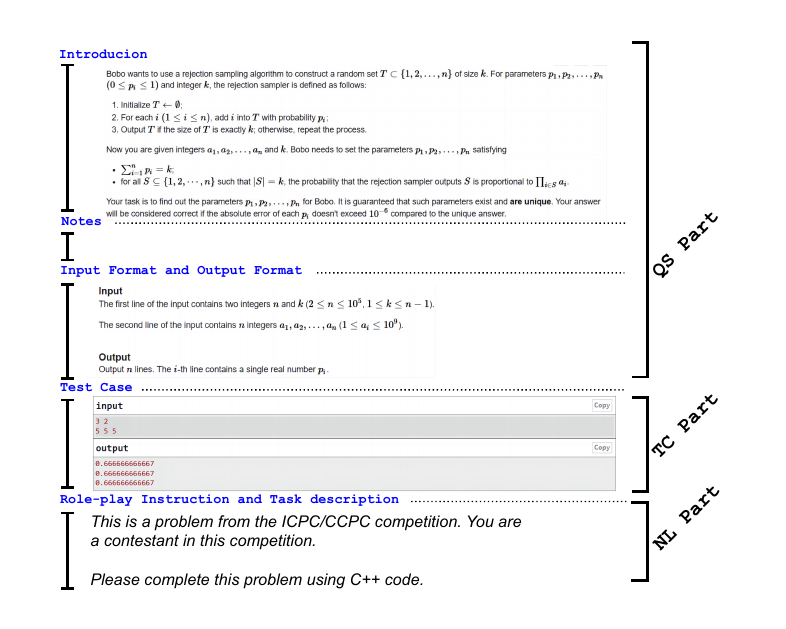}
    \caption{Basic prompt of 2024 China Collegiate Programming Contest (CCPC) Zhengzhou Onsite, problem M (Rolling Stones).} 
        \label{Fig:Improvingbasicprompt} 
    \end{figure}

\begin{figure*}[htbp] 
        \centering 
        \includegraphics[width=0.95\textwidth]{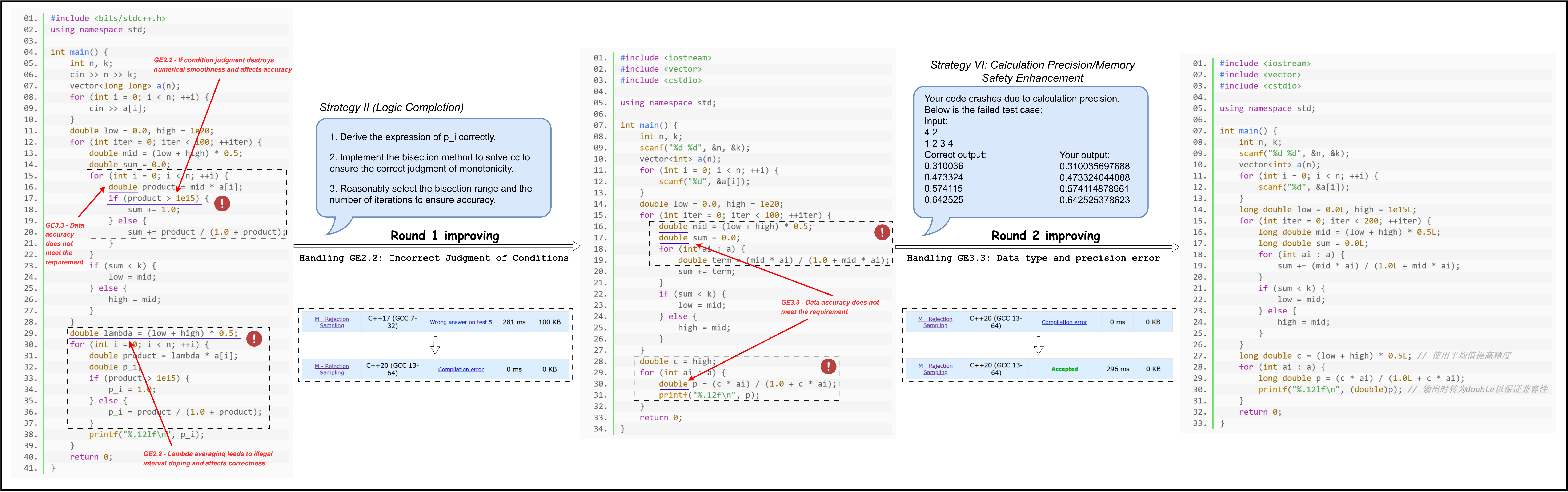}
    \caption{A multi-turn dialogue-based repair strategy example.} 
        \label{Fig:Improving} 
    \end{figure*}

An example (Problem M\footnote{\url{https://codeforces.com/gym/105632/problem/M}} from CCPC-ZZ) is used to demonstrate the repair process. The basic prompt for this example can be found in Figure~\ref{Fig:Improvingbasicprompt}, which serves as the foundation for subsequent multi-turn repair.
As shown in Figure~\ref{Fig:Improving}, the dialogue-based repair phrase progressively enhances the correctness of LLM-generated programs: it first targets higher-level logic errors (GE2.2 – incorrect condition judgements) through mathematical correction and algorithm-refinement techniques, and then resolves lower-level implementation issues (GE3.3 - data-type and precision errors) via precision-enhancement and memory-safety strategies. This staged workflow systematically improves both algorithm design and implementation quality, enabling the competitive program to evolve from an initial \texttt{WA5} status, to \texttt{CE} in the second round, and finally to \texttt{AC} after all test cases pass.

\subsection{Phase 3: Information-Augmented Regeneration.}  

Programs that remain incorrect after phase 1 are regenerated from scratch with a prompt template. Following Liu et al.~\cite{liu2023lost}, lengthy conversations can induce forgetting and hallucination; therefore, we inject structured, task-specific scaffolding (i.e., Problem Modeling, Algorithm Objectives, Algorithm Ideas, Implementation Details, and Constraints) into the new prompt (prompt template is shown on Github~\footnote{\url{https://github.com/minnanWei/LLMs-Competitive-Program-Generation/blob/main/README.md?plain=1}}). 
This holistic context enables the LLM to design a fresh, coherent solution instead of patching isolated defects.

\begin{figure}[htbp] 
    \centering 
    \includegraphics[width=0.45\textwidth]{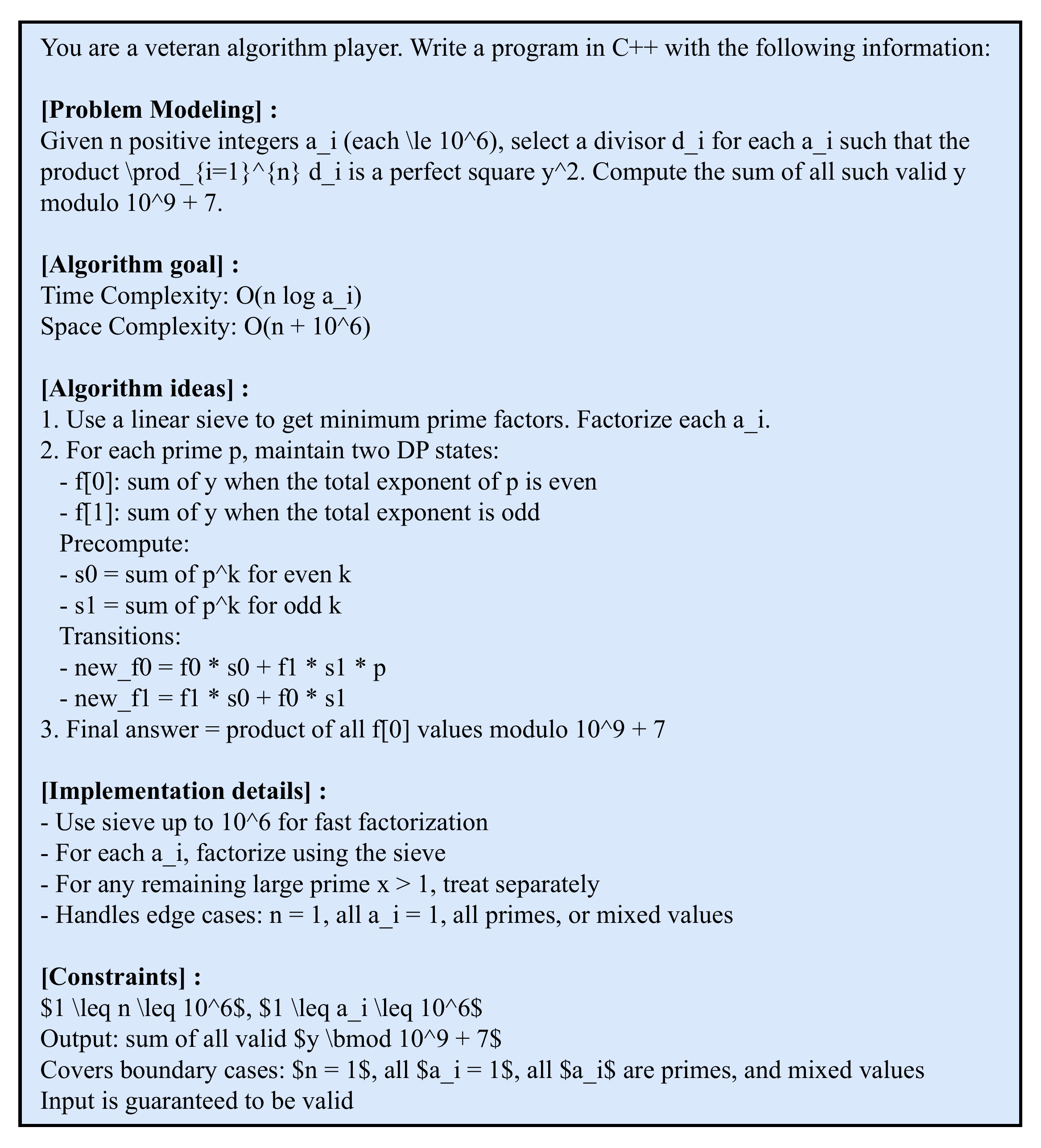}
\caption{A prompt template example for CCPC-FM, Problem F.} 
    \label{Fig:phase2} 
\end{figure}

Give an example to demonstrate the information-augmented regeneration prompt template: CCPC-FM, Problem F (Perfect Square)\footnote{\url{https://codeforces.com/gym/105487/problem/F}}, The program generated by the LLM based on the basic prompt produced an incorrect result on the first test case.  However, after applying the basic prompt combined with the information-augmented regeneration prompt template (shown in Figure~\ref{Fig:phase2}), the regenerated program successfully passed all test cases and achieved an AC state.

\section{Improvement Results}

\begin{figure}[htbp] 
    \centering 
    \includegraphics[scale=0.50]{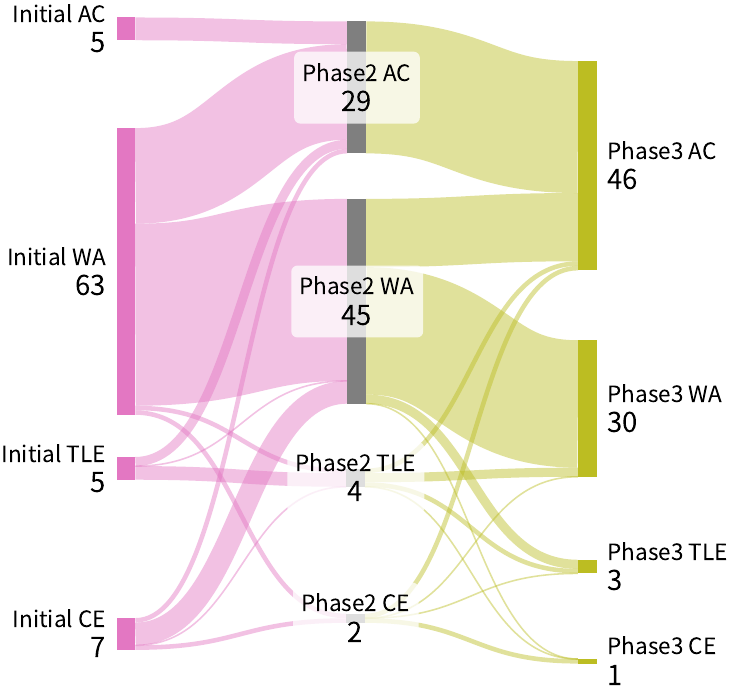}
\caption{Comparison of program generation results before and after improvement} 
    \label{fig:sankey-two-phase} 
\end{figure}

After applying our improvement framework, the number of fully correct (AC) solutions increases from 5 to 46 out of 80 problems, while significantly reducing incorrect outputs (i.e., WA decreased from 63 to 30, TLE from 5 to 3, and CE from 7 to 1). The complete transition of program statuses across the two phases is illustrated in the Sankey diagram, as shown in  Figure~\ref{fig:sankey-two-phase}. These results demonstrate the effectiveness of structured error categorization and targeted repair in enhancing LLM-based competitive programming capabilities. Notably, the proposed approach proves effective not only across different difficulty levels but also generalizes well to problems requiring either single or multiple algorithms. The improvement process followed a phased strategy: Phase 2 focused on correcting problems with relatively few errors and simple fixes (i.e., cases where the core logic was sound but minor implementation issues remained). In contrast, Phase 3 addressed more complex cases that Phase 2 could not resolve (i.e., programs with multiple interdependent errors or requiring substantial restructuring).

\begin{figure*}[htbp] 
    \centering 
    \includegraphics[width=0.95\textwidth]{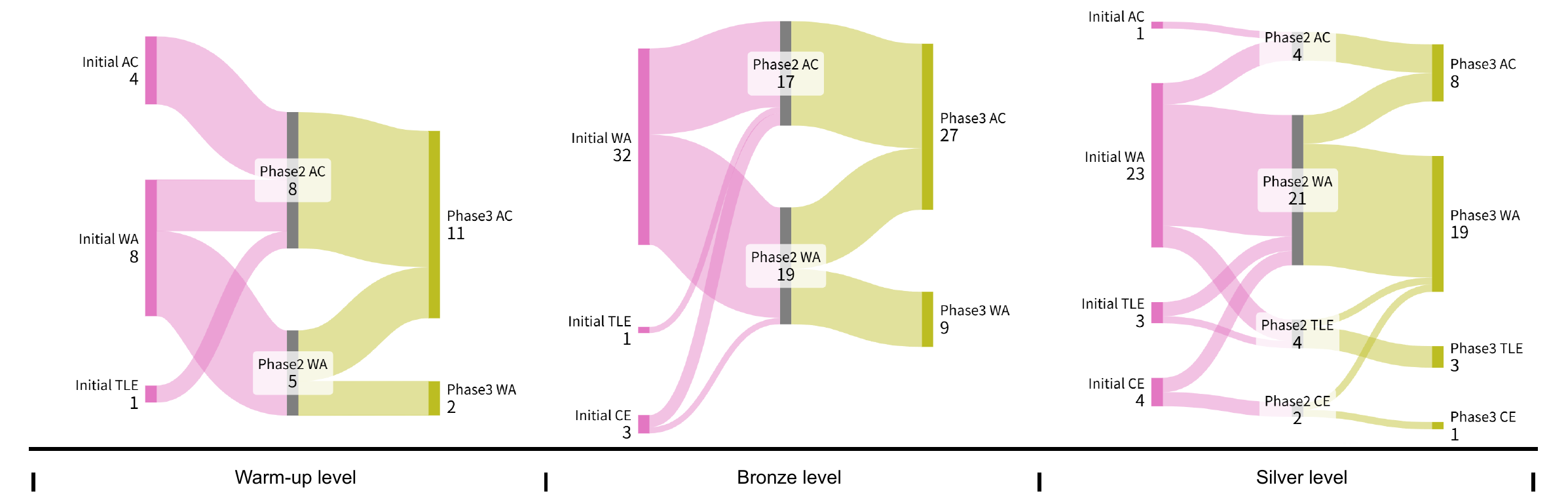}
\caption{Comparison of program generation results before and after improvement in terms of difficulty levels.} 
    \label{fig:analysisintermsofdifficultylevels} 
\end{figure*}

\textbf{\underline{Analysis in terms of difficulty levels}.}
Shown in Figure~\ref{fig:analysisintermsofdifficultylevels}, after applying the taxonomy-driven improvement framework, the generation results changed as follows. For warm-up problems, the results improved from 4 AC, 8 WA, 1 TLE under the basic prompt to 8 AC, 5 WA after Phase 2, and finally to 11 AC, 2 WA after Phase 3. For bronze-level problems, the outcomes progressed from 0 AC, 32 WA, 1 TLE, and 3 CE to 17 AC, 19 WA after Phase 2, and ultimately to 27 AC, 9 WA. For silver-level problems, the improvement trajectory was from 1 AC, 23 WA, 3 TLE, and 4 CE to 4 AC, 22 WA, 3 TLE, 2 CE after Phase 2, and then to 8 AC, 19 WA, 3 TLE, 1 CE after Phase 3.

\begin{figure}[htbp] 
    \centering 
    \includegraphics[width=0.45\textwidth]{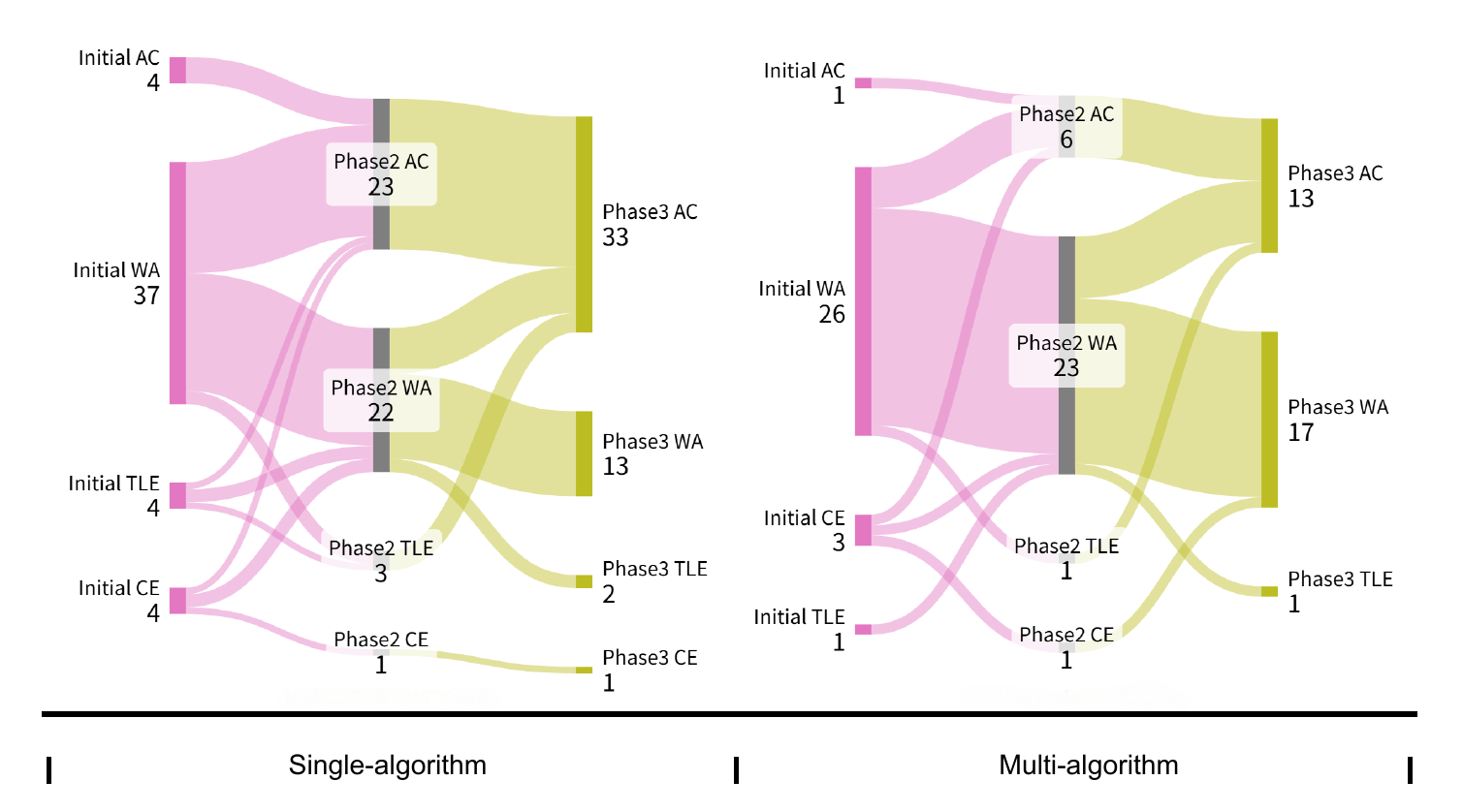}
\caption{Comparison of program generation results before and after improvement in terms of problem types} 
    \label{fig:analysisintermsoftheproblemtypes} 
\end{figure}

\textbf{\underline{Analysis in terms of the problem types.}}
Shown in Figure~\ref{fig:analysisintermsoftheproblemtypes}, after applying the taxonomy-driven improvement framework, the generation results also improved significantly in terms of problem types. For single-algorithm-type problems, the results improved from 4 AC, 37 WA, 4 TLE, and 4 CE under the basic prompt to 24 AC, 23 WA, 1 TLE, and 1 CE after Phase 2, and further to 33 AC, 13 WA, 2 TLE, and 1 CE after Phase 3. For multi-algorithm-type problems, the results progressed from 1 AC, 26 WA, 1 TLE, and 3 CE to 6 AC, 24 WA, 1 TLE after Phase 2, and ultimately to 13 AC, 17 WA, 1 TLE after Phase 3.

\section{Threats to Validity}
\label{sec:threatstovalidity}

\subsection{Internal Threats}

The first internal threat concerns the design of the basic prompts used in our evaluation. To mitigate this threat, we followed established practices from prior studies~\cite{reynolds2021prompt,beurer2023prompting,chen2023unleashing} and carefully crafted our basic prompts to align with the specific characteristics and requirements of competitive program generation tasks.


The second internal threat refers to the quality of the error taxonomy. Since the labeling process is performed manually, it can introduce subjectivity and inconsistency~\cite{yan2024errorradar}. To mitigate this risk, we followed established taxonomy construction methodologies~\cite{liu2024no}, where annotators independently labeled each instance. Disagreements were resolved through discussion, with a third party introduced to mediate and ensure consistency.

The final internal threat lies in the evaluation of program correctness. In our empirical study, we primarily rely on online judge platforms for validation. Typically, each problem is accompanied by comprehensive test cases, which not only cover normal functional scenarios but also include various boundary cases.

\subsection{External Threats}

The first external threat relates to the construction of our benchmark. To mitigate the risk of data leakage, we selected 80 problems from the 2024 ICPC and CCPC contests, which were released after the pre-training cutoff date of DeepSeek~\cite{zhang2024deep}. In addition, choosing problems from these contests helps ensure broader coverage across diverse problem types and difficulty levels.

The second external threat stems from the choice of the LLM. We selected DeepSeek-R1 due to its strong performance in code generation tasks and its accessibility as a state-of-the-art open-source model~\cite{shakya2025showdown}.

The final external threat concerns the choice of programming language during code generation. In our empirical study, we primarily focused on C++, as it is the dominant language in algorithmic competitions due to its high execution efficiency, rich standard library (e.g., STL), and widespread adoption in platforms like ICPC and Codeforces.

\section{Related Work}
\label{sec:relatedwork}

\textbf{Competitive program generation} focuses on automatically generating solutions to algorithmic problems from natural language descriptions. Compared with general-purpose coding tasks, this setting imposes stricter demands on algorithmic reasoning, correctness, and compliance with time/memory constraints and input/output specifications. As a result, it has become a critical benchmark for assessing the reasoning and planning abilities of LLMs.

\textbf{Benchmarks and datasets.} Early evaluations of program synthesis relied on datasets like HumanEval~\cite{chen2021evaluating} and MBPP~\cite{austin2021program}, which focus on small-scale functional problems and often lack coverage of complex algorithmic logic. More recent datasets like APPS~\cite{hendrycks2021measuring}, CodeContests~\cite{li2022competition} aim to better reflect competitive programming environments. However, many of these benchmarks suffer from data leakage due to overlap with training corpora, limited algorithm type diversity, and insufficient stratification across difficulty levels. This hinders the reliable evaluation of model generalization and robustness.

\textbf{Algorithmic code generation methods.} The release of \textbf{AlphaCode}\cite{li2022competition} marked a significant milestone by generating competitive programming solutions using a sampling-and-filtering pipeline. Follow-up work, such as \textbf{Self-Edit}\cite{zhang2023self} adopted runtime feedback and automatic correction to improve faulty outputs, while \textbf{AlphaCodium}\cite{ridnik2024code} proposed test-driven generation combined with iterative regeneration. Recent frameworks like \textbf{MapCoder}\cite{islam2024mapcoder} introduce a multi-agent structure for retrieval, planning, coding, and self-debugging. Lightweight models~\cite{souza2025code} and instruction-tuned systems like \textbf{Magicoder}\cite{lu2024magicoder} further explore efficiency and domain specialization. In terms of evaluation methodology, \textbf{ProBench}\cite{yang2025probench} presents an online-judge-based evaluation pipeline with fine-grained feedback, but lacks structured analysis of error types and model failure modes.

\textbf{Motivation and novelty of our study.} Despite recent progress, current datasets and evaluation settings still suffer from key limitations: (i)~potential data contamination due to overlaps with training data, (ii)~incomplete coverage of algorithm types (e.g., greedy, DP, graph), and (iii)~lack of stratification across difficulty levels. To address these issues, we construct a new benchmark by collecting and filtering problems from nine official ICPC and CCPC contests held in 2024. We then evaluate the latest LLM \textbf{DeepSeek-R1} on this benchmark using carefully designed basic prompts. The results show that only 5 out of 80 problems are fully solved. Motivated by this finding, we perform a systematic manual error analysis and construct a hierarchical error taxonomy to categorize common failure patterns. Finally, based on this taxonomy, we propose a \textbf{taxonomy-driven improvement framework} that combines multi-turn repair and information-augmented regeneration. Experiments show that this approach significantly improves solution correctness, increasing the number of accepted programs from 5 to 46, and providing concrete guidance for future work in LLM-based algorithmic code generation.

\section{Conclusion}
\label{sec:conclusion}

Our study addresses critical shortcomings in existing competitive programming datasets, such as data leakage and limited diversity in problem types and difficulty. To overcome these challenges, we developed a comprehensive benchmark sourced from recent ICPC and CCPC contests. Our evaluation of the DeepSeek-R1 model on this benchmark revealed significant performance limitations, motivating an in-depth error analysis and the creation of a detailed error taxonomy. Building upon these insights, we introduced an improvement framework that substantially enhances code generation accuracy across a broader spectrum of problems. These findings not only highlight current gaps in LLM capabilities but also offer a promising foundation for advancing future research in automated competitive programming code generation.

In future work, we will continue to expand our benchmark by incorporating problems from top-tier competitions such as Codeforces, the ICPC series, and the IOI series.
We also plan to evaluate more advanced LLMs, such as GPT-4, Claude, and Gemini. Furthermore, we aim to design and implement more effective optimization strategies specifically tailored to address the most challenging problems.

\section*{Acknowledegments}
This research was partially supported by the National Natural Science Foundation of China (Grant no. 61202006) and the Postgraduate Research \& Practice Innovation Program of Jiangsu Province (Grant no. SJCX24\_2022).

\section*{Declaration of Competing Interests}

The authors declare that they have no known competing financial interests or personal relationships that could have appeared to influence the work reported in this paper.

\section*{CRediT Authorship Contribution Statement}
\textbf{Minnan Wei:} Data curation, Writing -review \& editing, Software, Validation.
\textbf{Ziming Li:} Data curation, Software, Validation.
\textbf{Xiang Chen:} Software, Conceptualization, Methodology, Writing -review \& editing, Supervision.
\textbf{Menglin Zheng:} Data curation, Writing-review \& editing, Validation.
\textbf{Ziyan Qu:} Data curation, Software, Validation.
\textbf{Cheng Yu}: Data curation, Software, Validation.
\textbf{Siyu Chen}: Data curation, Software, Validation.
\textbf{Xiaolin Ju:} Methodology, Writing -review \& editing.

\bibliographystyle{elsarticle}
\bibliography{mylib}

\vspace{1cm}

\noindent\textbf{Minnan Wei} is currently pursuing a Master's degree at the School of Artificial Intelligence and Computer Science, Nantong University. His research interests include competitive program generation, 
 and vulnerability detection.

\par
\vspace{1cm}

\noindent\textbf{Ziming Li} is currently pursuing his Bachelor's degree at the School of Artificial Intelligence and Computer Science, Nantong University. 
His research interests include competitive program generation.
\par
\vspace{1cm}

\noindent\textbf{Xiang Chen} received the B.Sc. degree in information management and systems from Xi'an Jiaotong University, China, in 2002. Then he received his M.Sc. and Ph.D. degrees in computer software and theory from Nanjing University, China, in 2008 and 2011, respectively. He is currently an Associate Professor at the School of Artificial Intelligence and Computer Science, Nantong University. He has authored or co-authored more than 160 papers in refereed journals or conferences, such as IEEE Transactions on Software Engineering, ACM Transactions on Software Engineering and Methodology, Empirical Software Engineering, Information and Software Technology, Journal of Systems and Software, Software Testing, Verification and Reliability, Journal of Software: Evolution and Process, International Conference on Software Engineering (ICSE), International Conference on the Foundations of Software Engineering (FSE), International Symposium on Software Testing and Analysis (ISSTA), International Conference Automated Software Engineering (ASE), International Conference on Software Maintenance and Evolution (ICSME), International Conference on Program Comprehension (ICPC), and International Conference on Software Analysis, Evolution and Reengineering (SANER). His research interests include software engineering, in particular large language models for software engineering, software testing and maintenance, software repository mining, and empirical software engineering. He received two ACM SIGSOFT distinguished paper awards in ICSE 2021 and ICPC 2023. He is the editorial board member of Information and Software Technology. More information can be found at: 

https://xchencs.github.io/index.html.

\par
\vspace{1cm}

\noindent\textbf{Menglin Zheng} is currently pursuing a Bachelor’s degree in Software Engineering at the School of Artificial Intelligence and Computer Science, Nantong University, with a focus on artificial intelligence and multimodal analysis.

\par
\vspace{1cm}

\noindent\textbf{Ziyan Qu} is currently pursuing his Bachelor's degree at the School of Artificial Intelligence and Computer Science, Nantong University.
His research interests include competitive program generation.

\par
\vspace{1cm}

\noindent\textbf{Cheng Yu} is currently pursuing his Bachelor's degree at the School of Artificial Intelligence and Computer Science, Nantong University.
His research interests include competitive program generation.

\par
\vspace{1cm}

\noindent\textbf{Siyu Chen} is currently pursuing a Master's degree at the School of Artificial Intelligence and Computer Science, Nan-tong University. Her research interests include software vulnerability analysis.

\par
\vspace{1cm}

\noindent\textbf{Xiaolin Ju} was born in April 1976. He received the B.S. degree in information science from Wuhan University in 1998, the M.Sc. degree in computer science from Southeast University in 2004, and the Ph.D. degree in computer science from the Chinese University of Mining Technology in 2014. He is currently an associate professor at the School of Information Science and Technology, Nantong University, Nantong, China. His current research interests include software testing, such as collective intelligence, deep learning testing and optimization, and software defects analysis.


\end{document}